\documentclass[11pt,reqno,twoside]{article}
\usepackage{amsmath,amssymb,theorem,color,epsfig,epic,subfigure,amsfonts,hhline,graphicx}

\theorembodyfont{\itshape}
\newtheorem{thm}{Theorem}[section]
\newtheorem{lem}[thm]{Lemma}
\newtheorem{prop}[thm]{Proposition}
{\theorembodyfont{\upshape}
\newtheorem{define}[thm]{Definition}
\newtheorem{rem}[thm]{Remark}
\newtheorem{ex}[thm]{Example}
}
\newtheorem{cor}[thm]{Corollary}

\newcommand{\Proof}[1][]{\noindent{\itshape Proof#1. }}
\newcommand{\EndProof}{~$\Box$\bigskip}
\makeatletter
    
    \@addtoreset{equation}{section}
\makeatother

\def\hat{\widehat}

\def\mbR{{\mathbb R}}

\def\mcP{\mathcal{P}}

\def\sbs{\subset}

\def\wt{\widetilde}
\def\wh{\widehat}

\def\d{\delta}    
\def\a{\alpha}                  \def\d{\delta}
            \def\Ph{\Phi}
             
        \def\m{\mu}        \def\n{\nu}
          \def\o{\omega}     \def\O{\Omega}
\def\p{\pi}                   
\def\th{\theta}

\begin{document}
\title{Multivariate orthogonal polynomials: quantum decomposition, deficiency rank and support of measure}
\author{Ameur Dhahri\footnote{Department of Mathematics, Chungbuk National University, 1 Chungdae-ro, Seowon-gu, Cheongju, Chungbuk 28644, Korea. E-mail: ameur@chungbuk.ac.kr}, Nobuaki Obata\footnote{Graduate School of Information Sciences, Tohoku University, Sendai, 980-8579, Japan. E-mail: obata@tohoku.ac.jp}, and
   Hyun Jae Yoo\footnote{Department of Applied Mathematics,
Hankyong National University, 327 Jungang-ro, Anseong-si,
Gyeonggi-do 17579, Korea. E-mail: yoohj@hknu.ac.kr} \footnote{Corresponding author}}
\date{}
   \maketitle

\begin{abstract}
In this paper we investigate the multivariate orthogonal polynomials based on the theory of interacting Fock spaces. Our framework is on the same stream line of the recent paper by Accardi, Barhoumi, and Dhahri \cite{ABD}. The (classical) coordinate variables are decomposed into non-commuting (quantum) operators called creation, annihilation, and preservation operators, in the interacting Fock spaces. Getting the commutation relations, which follow from the commuting property of the coordinate variables between themselves, we can develop the reconstruction theory of the measure, namely the Favard's theorem. We then further develop some related problems including the marginal distributions and the rank theory of the Jacobi operators. We will see that the deficiency rank of the Jacobi operator implies that the underlying measure is supported on some algebraic surface and vice versa. We will provide with some examples.
\end{abstract}
\noindent {\bf Keywords}. {Multivariate orthogonal polynomials, quantum decomposition, Favard theorem, deficiency rank, support of measure.}  

\section{Introduction}\label{sec:introduction} 

The aim of this paper is to develop the study of
multivariate orthogonal polynomials within the 
formalism of interacting Fock spaces.
 
The univariate case has been studied in terms of one-mode
interacting Fock spaces, where the Jacobi's three-term recurrence
relation is transformed into annihilation, creation
and preservation operators through the quantum decomposition of
the multiplication operator by $x$.
This aspect traces back to Accardi and Bo\.zejko \cite{AB},
and is now well understood with many applications, see e.g.,
\cite{HO} and references cited therein.

On the other hand, multivariate case has been also
formulated within multi-mode interacting Fock spaces,
where the coordinate variables are decomposed into a sum of 
creation, annihilation, and preservation operators 
in the interacting Fock space of the gradations of polynomials. 
In that case the Jacobi coefficients appearing 
in the three-term recurrence relation should be 
replaced with a pair of sequences of
positive definite matrices and Hermitian ones
\cite{ABD, AKS, AN}. 
Following the formulation established therein,
we study multivariate version of Favard's theorem,
and discuss the relation between 
the support of the probability measure 
and the Jacobi coefficients,
where we use a newly introduced concept of deficiency rank.

Given a probability measure on $\mathbb R^d$ with finite moments of
all orders, we perform the Gram-Schmidt orthogonalization process
and obtain the gradation spaces (spaces of polynomials of degree $n$ for each $n\ge 0$ in the orthogonalization process). 
As is the case of univariate system, the coordinate variables
$x,y,\dots$ are decomposed into the three (non-commuting) operators between gradation spaces, which are called creation, annihilation, and preservation (or conservation) operators (called CAPs, hereafter). 
The commutativity of the (classical) variables themselves require some commutation relations for the CAPs. In view of this structure, we next consider the converse problem, namely the Favard's theory. Starting with an interacting Fock space provided with CAP operators which satisfy suitable conditions, we reconstruct a probability measure. 
For this we will use the spectral theory of mutually commuting operators after Xu \cite{Xu1, Xu2}.
We will see that the commutation relations are so strong that already for the simplest case of product measures, they must obey some rules to properly construct the measure. Moreover,
we introduce a sequence $\{\rho_n=\mathrm{rank\,} \Omega_n\}$ 
of the ranks of the Jacobi operators (matrices)
and investigate the relation to the support of the measures. 
It would be interesting question to characterize
probability measures in terms of the rank sequence $\{\rho_n\}$.

There are tremendous works on multivariate orthogonal
polynomials from various aspects,
see \cite{DX, K1, K2, KS, Su, Xu1, Xu2, Xu3, Xu4, Xu5}
and references cited therein.
Our approach has an algebraic feature that enables us to
use commutation relations of CAP operators and to discuss
the supports of probability measures.

The organization of the paper is as follows. In Section 2, we shortly recall the univariate theory of orthogonal polynomials.  In Section 3, we develop the interacting Fock spaces for the multivariate orthogonal  polynomials by using CAPs. Section 4 deals with the reconstruction theory, which is a multivariate version of Favard's theorem. In Section 5, we introduce the form generators for the gradation spaces in the interacting Fock space. Section 6 is devoted to some examples. In Section 7 we deal with the marginals. In Section 8, we introduce the concept of deficiency rank of Jacobi operators and discuss the support of the measure. 
 
\section{Preliminary}\label{sec:preliminary}

In this section we briefly review the theory of univariate orthogonal polynomials. In this paper, by a measure on $\mathbb R^d$ we mean a Borel measure without specifying the Borel $\sigma$-field anymore. Let $\mu$ be a probability measure on $\mathbb R$ such that the moments of all orders exist. Let $\{p_n(x)\}$ be the monic orthogonal polynomials of $\mu$ obtained by Gram-Schmidt orthogonalization of $\{1,x,x^2,\cdots\}$. Then there exist Jacobi sequences $\{\omega_n\}_{n\ge 1}$ and $\{\alpha_n\}_{n\ge 1}$ such that the three-term recurrence relation holds:
\begin{eqnarray}\label{eq:three-term_recurrence}
p_0(x)&=&1,\nonumber\\
p_1(x)&=&x-\alpha_1,\nonumber\\
xp_n(x)&=&p_{n+1}(x)+\alpha_{n+1}p_n(x)+\omega_np_{n-1}(x), \quad n\ge 1.
\end{eqnarray}
Here we notice that  $\omega_n>0$ for all $n\ge 1$ or there exists $m_0\ge 1$ such that $\omega_n=0$ for all $n\ge m_0$ and $\omega_n>0$ for all $n<m_0$, and $\alpha_n\in \mathbb R$, $n\ge 1$ \cite{HO}. 

The Favard theorem says the converse: if there are Jacobi sequences $\{\omega_n\}_{n\ge 1}$ and $\{\alpha_n\}_{n\ge 1}$, then there is a probability measure on $\mbR$ for which the polynomials $\{p_n(x)\}$, constructed via the three-term recurrence relation \eqref{eq:three-term_recurrence}, are orthogonal.  

Orthogonal polynomials can also be understood by using an interacting Fock space and  CAP operators \cite{AB}. Let $\mathcal H$ be the direct sum Hilbert space:
\[
\mathcal H:=\oplus_{n=0}^\infty \mathbb C^{\widehat\otimes n}.
\]
For $n\ge 0$, let $\Phi_n:={\bf e}_1^{\widehat\otimes n}$, where ${\bf e}_1:=1\in \mathbb C$.  
Given a Jacobi sequence $(\{\omega_n\},\{\a_{n}\})$, define linear operators $A^+$, $A^-$, and $A^0$ on $\mathcal H$ by 
\begin{eqnarray}\label{eq:1-d_CAP}
\label{eq:1-d_C} A^+\Phi_n&=&\sqrt{\omega_{n+1}}\Phi_{n+1}, \quad n\ge 0,\\
\label{eq:1-d_P} A^0\Phi_n&=&\alpha_{n+1}\Phi_n, \quad n\ge 0,\\
\label{eq:1-d_A} A^-\Phi_n&=&\sqrt{\omega_n}\Phi_{n-1}, \quad n\ge 1,\quad A^-\Phi_0=0.
\end{eqnarray}
The Jacobi coefficients, orthogonal polynomials, and interacting Fock space and the CAP operators have the following relations \cite{HO}:
\begin{thm}\label{eq:1-d_IFS}
Let $(\{\omega_n\}, \{\alpha_n\})$ be the Jacobi coefficients for a probability measure $\mu$ on $\mathbb R$ having all moments of any order and let  $\{p_n(x)\}$ the corresponding monic orthogonal polynomials. Let $\mathcal H$ be the interacting Fock space with CAP operators in \eqref{eq:1-d_C}-\eqref{eq:1-d_A} on it. Then the map 
\[
U:\mathcal H\ni \Phi_n\mapsto (\omega_n\cdots \omega_1)^{-1/2}p_n\in L^2(\mathbb R,\mu),\quad n=1,2,\cdots,
\]
and defined by linear extension, is a unitary operator. It holds also that the multiplication operator by $x$ on $L^2(\mathbb R,\mu)$, denoted by $X$ has a representation:
\[
X=U(A^++A^0+A^-)U^*,
\]
which we call a quantum decomposition. Furthermore, the following relation for the moments holds:
\begin{equation}\label{eq:1-dim_moments_formula}
\int_{-\infty}^\infty x^md\mu(x)=\langle \Phi_0,(A^++A^0+A^-)^m\Phi_0\rangle_{\mathcal H}, \quad m=0,1,2,\cdots.
\end{equation}
\end{thm}
In this paper we extend the theory to the multivariate functions. 

\section{Interacting Fock spaces}\label{sec:multi-dim_IFS}
In this section, given a probability measure on $\mathbb R^d$ with finite moments of any order, we introduce an interacting Fock space and CAP operators. Then we represent the moments via vacuum expectation, which is definitely an extension of one-dimensional theory. All the basic ideas are already given in the reference \cite{ABD}, but here we deal with in a canonical setting and this will make the argument more clear.  

Throughout this section, we fix a probability measure $\mu$ on $\mathbb R^d$ such that the moments of $\mu$ of any order are finite. For a basic setting we follow \cite{ABD}: we define \lq\lq gradation\rq\rq spaces. Let $\mathcal P_{n]}$ be the space of all polynomials of degree $n$. 
Here we emphasize that the spaces $\mathcal P_{n]}$, $n\ge 0$, are understood as pre-Hilbert spaces equipped with a pre-scalar product $\langle \cdot,\cdot\rangle_\mu$, the $L^2$-inner product. Thus any two polynomials $f$ and $g$ are equivalent, or regarded as the same vector if $\int|f-g|^2d\mu=0$, i.e., $f=g$ $\mu$-a.e. Having this equivalence relation in mind we may think of $\mathcal P_{n]}$'s as Hilbertian subspaces of $L^2(\mathbb R^d,\mu)$.

 We call $\mathcal P_n:=\mathcal P_{n]}\ominus \mathcal P_{n-1]}$, $n=0,1,2,\cdots$, the $n$th gradation space ($\mathcal P_{-1]}:=\{0\}$). In other words, $\mathcal P_n$ consists of polynomials of degree $n$ subtracted by their orthogonal projections onto $\mathcal P_{n-1]}$. We therefore get the following direct sum structure:
\[
\mathcal P_{n]}=\oplus_{k=0}^n\mathcal P_k, \quad n=0,1,2,\cdots.
\]
From now on, the projection operators onto the spaces $\mathcal P_n$'s will be denoted by $P_n$'s and similarly by $P_{n]}$'s for the projection operators onto $\mathcal P_{n]}$'s. We let $\mathcal P$ the space of all polynomials. Notice that $\mathcal P\subset \mathcal K:=\oplus_{n=0}^\infty\mathcal P_n$. Below the constant unit function $1\in \mathcal P$ is explicitly exposed whenever some operation is done on it.

For each $i=1,\cdots,d$, we understand the variable $x_i$ also as a multiplication operator by $x_i$ defined on $\mathcal P$. We notice here that when we consider $x_i$ as an operator on $L^2(\mathbb R^d,\mu)$, it is an unbounded operator unless $\mu$ is compact supported. In that case we take $\mathcal P$ as the space of definition which is dense in $\mathcal K=\oplus_{n=0}^\infty\mathcal P_n$. We remark that $\mathcal K$ may not be equal to $L^2(\mathbb R^d, \mu)$. 

It is easy to check and has been shown in \cite[Theorem 4.2]{ABD} that 
\begin{equation}\label{eq:multiplication_decomposed}
x_i  P_n=  P_{n+1}x_i  P_n+  P_{n}x_i  P_n+  P_{n-1}x_i  P_n.
\end{equation}
We can thus define creation, preservation, and annihilation operators on $\mathcal P$, denoted by $a_i^+$, $a_i^0$, $a_i^-$, $i=1,\cdots,d$, in that order,  as follows.
\begin{eqnarray}
\label{eq:C_polynomial} \left. a_i^+\right|_{\mathcal P_n}&:=&  P_{n+1}x_i  P_n,\\
\label{eq:P_polynomial} \left. a_i^0\right|_{\mathcal P_n}&:=&  P_{n}x_i  P_n,\\
\label{eq:A_polynomial} \left. a_i^-\right|_{\mathcal P_n}&:=&  P_{n-1}x_i  P_n, \quad \left. a_i^-\right|_{\mathcal P_0}:=0.
\end{eqnarray}
Therefore we have the following relation, called quantum decomposition.
\begin{equation}\label{eq:quantum_decomposition}
x_i=a_i^++a_i^0+a_i^-, \quad i=1,\cdots,d, \text{ on }\mathcal P.
\end{equation}
As was shown in \cite{ABD}, we notice that $\{a_i^+:i=1,\cdots,d\}$ is a set of mutually commuting operators. Moreover, $\left.(a_i^+)^*\right|_{\mathcal P}=a_i^-$, and $a_i^0$ is a symmetric operator for each $i=1,\cdots,d$.

Now we transfer the story into the canonical interacting Fock space over $\mathbb C^d$. For each $n=0,1,2,\cdots$, we define the set of multi-indices:
\begin{equation}\label{eq:multi_index}
\mathcal I^{(n)}\equiv\mathcal I_d^{(n)}:=\{{\bf n}:=(n_1,\cdots,n_d):n_i\ge 0, \,i=1,\cdots, d,\,|{\bf n}|:=n_1+\cdots+n_d=n\}.
\end{equation} 
Let $\mathcal H_0:=\mathbb C$ and for each $n\ge 1$ let $\mathcal H_n$ be the vector space $(\mathbb C^d)^{\widehat\otimes n}$ equipped with a pre-scalar product $\langle \cdot,\cdot\rangle_n$ defined as follows: for ${\bf n}=(n_1,\cdots,n_d), \,{\bf m}=(m_1,\cdots,m_d)\in \mathcal I^{(n)}$, 
\begin{equation}\label{eq:inner_product}
\langle {\bf e}_1^{\widehat\otimes n_1}\widehat \otimes \cdots \widehat \otimes {\bf e}_d^{\widehat\otimes n_d},{\bf e}_1^{\widehat\otimes m_1}\widehat \otimes \cdots \widehat \otimes {\bf e}_d^{\widehat\otimes m_d}  \rangle_n :=\langle(a_1^+)^{n_1}\cdots(a_d^+)^{n_d}1,
(a_1^+)^{n_1}\cdots(a_d^+)^{n_d}1\rangle_{\mu} .
\end{equation}
We identify $\mathcal H_0\equiv \mathbb C {\Ph_0}$, where ${\Ph_0}$ is any fixed unit vector (a symbol), called vacuum vector. Notice that $\{{\bf e}_1^{\widehat\otimes n_1}\widehat \otimes \cdots \widehat \otimes {\bf e}_d^{\widehat\otimes n_d}:\,{\bf n}=(n_1,\cdots,n_d)\in \mathcal I^{(n)}\}$ is an (not normalized) orthogonal basis for $(\mathbb C^d)^{\widehat \otimes n}$ with the canonical inner product, but it is not an orthogonal system for $\mathcal H_n$ in general. For each $n\ge 0$, we define a linear operator $U_n:\mathcal H_n\to \mathcal P_n$ by 
\begin{equation}\label{eq:unitary}
U_n({\bf e}_1^{\widehat\otimes n_1}\widehat \otimes \cdots \widehat \otimes {\bf e}_d^{\widehat\otimes n_d}) :=(a_1^+)^{n_1}\cdots(a_d^+)^{n_d}1, 
\end{equation} 
and by a linear extension. We easily check that $U_n$ is an isomorphic unitary. We define an interacting Fock space:
\begin{equation}\label{eq:IFS}
\mathcal H:=\oplus_{n=0}^\infty\mathcal H_n.
\end{equation}
By defining $U:=\oplus_{n=0}^\infty U_n$, the operator $U:\mathcal H\to \mathcal K$ becomes again an isomorphic unitary. We transfer the CAP operators into $\mathcal H$ by 
\begin{equation}\label{eq:CAP_canonical}
A_i^+:=U^*a_i^+U, \quad A_i^0:=U^*a_i^0U, \quad A_i^-:=U^*a_i^-U, \quad i=1,\cdots,d.
\end{equation}
Since the domain of CAP operators $\{a_i^+,a_i^0,a_i^-:i=1,\cdots,d\}$ as well as $x_i$, $i=1,\cdots, d$, are $\mathcal P$, the domain for the CAP operators $\{A_i^+,A_i^0,A_i^-:i=1,\cdots,d\}$ are $\mathcal D:=U^{-1}\mathcal P$. Notice that any  element $\sum_{n=0}^\infty\xi_n\in \mathcal D$ with $\xi_n\in \mathcal H_n$ has at most finitely many non-zero terms $\xi_n$.

By definition the creation operator $A_i^+$ has always a canonical form in the sense that   for ${\bf n}=(n_1,\cdots,n_d)\in \mathcal I^{(n)}$,
\begin{equation}\label{eq:creation}
 \left. A_i^+\right|_{\mathcal H_n}({\bf e}_1^{\widehat\otimes n_1}\widehat \otimes \cdots \widehat \otimes {\bf e}_d^{\widehat\otimes n_d})={\bf e}_1^{\widehat\otimes n_1}\widehat \otimes \cdots \widehat \otimes {\bf e}_i^{\widehat\otimes (n_i+1)} \widehat \otimes \cdots \widehat \otimes {\bf e}_d^{\widehat\otimes n_d} .
\end{equation}

The set of CAP operators $\{A_i^+,A_i^0,A_i^-:i=1,\cdots,d\}$ inherits the properties from the set of CAP operators $\{a_i^+,a_i^0,a_i^-:i=1,\cdots,d\}$. In particular we see that $\{A_i^+:i=1,\cdots,d\}$ is a set of mutually commuting operators, $\left.(A_i^+)^*\right|_\mathcal D=A_i^-$, and $A_i^0$ is a symmetric operator for each $i=1,\cdots,d$. Let us define
\begin{equation}\label{eq:multiplication_operator_IFS}
X_i:=A_i^++A_i^0+A_i^-, \quad i=1,\cdots,d, \text{ on }\mathcal D.
\end{equation}
Then 
\begin{equation}\label{eq:multiplication_canonical}
X_i=U^*x_iU, \quad i=1,\cdots,d.
\end{equation}
From the commutativity of $\{x_i:i=1,\cdots,d\}$, we see that $\{X_i:i=1,\cdots,d\}$ is a set of commuting operators.   Moreover, the following commutation relations hold on the domain $\mathcal D$ (see \cite{ABD}): for all $j,k=1,\cdots,d$,
\begin{eqnarray}
\label{eq:CR_1}[A_j^+,A_k^+]&=&0,\\ 
\label{eq:CR_2}{[A_j^+,A_k^0]}+[A_j^0,A_k^+]&=&0,\\ 
\label{eq:CR_3}{[A_j^+,A_k^-]}+[A_j^0,A_k^0]+[A_j^-,A_k^+]&=&0.\label{eq:CR_3}
\end{eqnarray}
Taking adjoint, the relation \eqref{eq:CR_2} is equivalent to ${[A_j^0,A_k^-]}+[A_j^-,A_k^0]=0$.  

Now we have interacting Fock space $\mathcal H$ and the creation, annihilation, and preservation operators. It is then possible to compute the mixed moments of $\mu$ by the vacuum expectation.
\begin{prop}\label{prop:moments}
For any ${\bf n}=(n_1,\cdots,n_d)\in \mathcal I^{(n)}$, we have
\begin{eqnarray}\label{eq:multi-dim_vacuum_expectation}
\int_{\mathbb R^d}x_1^{n_1}\cdots x_d^{n_d}d\mu&=&\langle {\Ph_0},X_1^{n_1}\cdots X_d^{n_d}{\Ph_0}\rangle _0\nonumber\\
&=&\langle {\Ph_0},(A_1^++A_1^0+A_1^-)^{n_1}\cdots (A_d^++A_d^0+A_d^-)^{n_d}{\Ph_0}\rangle _0.
\end{eqnarray}
\end{prop}
\Proof
We notice that 
\[
\int_{\mathbb R^d}x_1^{n_1}\cdots x_d^{n_d}d\mu=\langle 1,x_1^{n_1}\cdots x_d^{n_d}1\rangle_\mu.
\] 
The result now follows from the relation \eqref{eq:multiplication_canonical} and the fact that $U:\mathcal H\to \mathcal K$ is an isomorphic unitary. 
\EndProof
\begin{rem}\label{rem:computation}
The relation \eqref{eq:multi-dim_vacuum_expectation} is an extension of the univariate formula \eqref{eq:1-dim_moments_formula}. By expansion, the r.h.s. of \eqref{eq:multi-dim_vacuum_expectation} is a linear combination of the terms:
\[
\langle {\Ph_0}, A_1^{\epsilon_{1,1}}\cdots A_1^{\epsilon_{1,n_1}}\cdots A_d^{\epsilon_{d,1}}\cdots A_d^{\epsilon_{d,n_d}}{\Ph_0}\rangle _0, \quad \epsilon_{i,k}\in \{+,0,-\}.
\] 
It is clear that each term with $\epsilon_{1,1}+\cdots+\epsilon_{1,n_1}+\cdots+\epsilon_{d,1}+\cdots+ \epsilon_{d,n_d}\neq 0$ is zero.
\end{rem}
An example will be discussed in subsection \ref{subsec:moments_uniform_circle}.

\section{Reconstruction theorem}\label{sec:reconstruction_theorem}
In this section, we discuss the converse problem. That is, given an interacting Fock space over $\mathbb C^d$ equipped with CAP operators we discuss how we can construct a probability measure on $\mathbb R^d$ so that its interacting Fock space structure is the given one. From the discussion of the previous section, it is clear  what kind of ingredients we have to have at hand a priori. Suppose that we are given an interacting Fock space
\begin{equation}\label{eq:IFS_for_measure}
\mathcal H:=\oplus_{n=0}^\infty \mathcal H_n,
\end{equation}
where $\mathcal H_n$ is the vector space $(\mathbb C^d)^{\widehat \otimes n}$ equipped with a pre-scalar product $\langle \cdot, \cdot \rangle_n$.   For $n\ge 0$, let $P_n$ be the orthogonal projection onto $n$th component space, $\cdots\oplus\{0\}\oplus\mathcal H_n\oplus\{0\}\oplus+\cdots$. We let $P_{n]}:=\sum_{k=0}^nP_k$. The creation operators $A_i^+:\mathcal H  \to \mathcal H$, $i=1,\cdots,d$, with a dense domain $\mathcal D$ which is a subspace of $\mathcal H$ consisting of finitely many components, are defined as in \eqref{eq:creation}, and we let $A_i^-$, $i=1,\cdots, d$, be the adjoints of $A_i^+$, $i=1,\cdots,d$, respectively, restricted on $\mathcal D$, i.e.,  $\left. A_i^-\right|_{\mathcal H_{n+1}}:\mathcal H_{n+1}\to \mathcal H_n$ is the adjoint of $\left.A_i^+\right|_{\mathcal H_{n}}\mathcal H_n\to \mathcal H_{n+1}$, they are called the annihilation operators. By convention we let $A_i^-{\Phi_0}:=0$. 
Suppose that for $i=1,\cdots,d$, we are also given preservation operators $A_i^0:\mathcal H\to \mathcal H$, which are symmetric operators and $\left. A_i^0\right|_{\mathcal H_n}:\mathcal H_n\to \mathcal H_n$. They may be all zero operators. On the dense subspace $\mathcal D$ let us define the following operators
\begin{equation}\label{eq:multiplication_IFS_reconstruct}
 X_i:=A_i^++A_i^0+A_i^-, \quad i=1,\cdots,d.
 \end{equation}
We notice that $X_i$'s are symmetric operators on $\mathcal D$. 
\begin{thm}\label{thm:reconstruction}
Suppose that there is a symmetric interacting Fock space over $\mathbb C^d$ and creation, annihilation, and preservation operators described above. Suppose that the operators $A_i^+$, $A_i^0$, and $A_i^-$, $i=1,\cdots, d$, satisfy the following conditions:
\begin{enumerate}
\item[(i)] $\|\left. A_i^+\right|_{\mathcal H_n}(\xi_n)\|_{n+1}=0$ and $\|\left. A_i^0\right|_{\mathcal H_n}(\xi_n)\|_{n}=0$ whenever $\|\xi_n\|_n=0$;\\[-3ex]
\item[(ii)] The commutation relations in \eqref{eq:CR_1}-\eqref{eq:CR_3} hold;\\[-3ex]
\item[(iii)] The symmetric operators $\{X_i:i=1,\cdots,d\}$ are essentially self-adjoint. Moreover, the closures $\{\overline{X}_i:i=1,\cdots,d\}$ are mutually commuting (in the sense that their spectral measures commute).
\end{enumerate} 
Then there is a probability measure $\mu$ on $\mathbb R^d$ such that its interacting Fock space constructed by the method in section \ref{sec:multi-dim_IFS} is the same as the one given a priori.
 \end{thm} 
\begin{rem}\label{rem:sufficiency_remark}
If the operators $\{X_i:i=1,\cdots,d\}$ are bounded, then the condition (iii) is automatically satisfied by Lemma \ref{lem:commuting_operators} below and the fact that commuting in spectral measures is equivalent to commuting in operator themselves for bounded operators. In the case that they are not bounded a sufficient condition for (iii) will be given in Proposition \ref{prop:sufficiency_for_(iii)}.
\end{rem}
The basic ingredients for the proof are the spectral theorem for commuting operators. This is also the main method used in \cite{Xu1,Xu2}. Notice that the condition (iii) is automatically satisfied if $X_i$'s are bounded, so it is needed when we deal with unbounded $X_i$'s. Before going further we prepare some basic properties. 
\begin{lem}\label{lem:commuting_operators}
The operators $X_i$, $i=1,\cdots,d$, are mutually commuting operators on $\mathcal D$:
 \[
 [X_j,X_k]=0, \quad j,k=1,\cdots,d.
 \]
\end{lem}
\Proof It follows directly from the commutation relations \eqref{eq:CR_1}-\eqref{eq:CR_3}.
\EndProof
\begin{lem}\label{lem:commutative_algebra}
Under the conditions (i) and (ii) of Theorem \ref{thm:reconstruction}, $\Phi_0$ is a cyclic vector w.r.t. $\{X_1,\cdots,X_d\}$ on   the symmetric interacting Fock space $\mathcal H=\oplus_{n=0}^\infty \mathcal H_n$. In particular for any ${\bf n}=(n_1,\cdots,n_d)\in \mathcal I^{(n)}$, we have the equality 
\begin{equation}\label{eq:cyclicity}
(A_1^+)^{n_1}\cdots(A_d^+)^{n_d}{\Phi_0}=P(X_1,\cdots,X_d){\Phi_0},
\end{equation}
where $P(x_1,\cdots,x_d)$ is a polynomial of degree $n$. 
\end{lem}  
\Proof
Since any element of $\mathcal D$ is a linear combination of the vectors $(A_1^+)^{n_1}\cdots(A_d^+)^{n_d}{\Phi_0}$ it is enough to prove the relation \eqref{eq:cyclicity}. For the proof we use an induction. For any $i=1,\cdots,d$, we see from \eqref{eq:multiplication_IFS_reconstruct} and the fact $A_i^-{\Phi_0}=0$ that
 \begin{eqnarray*}
 A_i^+{\Phi_0}&=&(X_i-A_i^0){\Phi_0}\\
 &=&(X_i-a_i^0){\Phi_0},
 \end{eqnarray*} 
 where $a_i^0\in \mathbb R$ is the matrix component of the one-dimensional linear operator $\left. A_i^0\right|_{\mathcal H_0}$. Suppose now that for any ${\bf k}=(k_1,\cdots,k_d)\in \mathcal I^{(k)}$ for $k\le n$, the claim holds:
 \[
(A_1^+)^{k_1}\cdots(A_d^+)^{k_d}{\Phi_0}=R(X_1,\cdots,X_d){\Phi_0},
\]
for some polynomial $R$ of degree $k$. Now let ${\bf n}=(n_1,\cdots, n_d)\in \mathcal I^{(n)}$. Then for any $i=1,\cdots,d$, 
\[
A_i^+\left((A_1^+)^{n_1}\cdots(A_d^+)^{n_d}\right){\Phi_0}=(X_i-A_i^0-A_i^-)\left((A_1^+)^{n_1}\cdots(A_d^+)^{n_d}\right){\Phi_0}.
\] 
For the second term in the r.h.s., $A_i^0 \left((A_1^+)^{n_1}\cdots(A_d^+)^{n_d}\right){\Phi_0}$, by definition of the operator $A_i^0$, it is a linear combination of the vectors $(A_1^+)^{m_1}\cdots(A_d^+)^{m_d}{\Phi_0}$, ${\bf m}=(m_1,\cdots,m_d)\in \mathcal I^{(n)}$. Similarly by definition of $A_i^-$, the third term $A_i^-\left((A_1^+)^{n_1}\cdots(A_d^+)^{n_d}\right){\Phi_0}$ is a linear combination of the vectors $(A_1^+)^{l_1}\cdots(A_d^+)^{l_d}{\Phi_0}$, ${\bf l}=(l_1,\cdots,l_d)\in \mathcal I^{(n-1)}$. By the induction hypothesis, the sum of those two terms is of the form $Q(X_1,\cdots, X_d){\Phi_0}$ for some polynomial of degree $n$. The first term is obviously of the form $X_iP(X_1,\cdots,X_d){\Phi_0}$, where we assumed $(A_1^+)^{n_1}\cdots(A_d^+)^{n_d}{\Phi_0}=P(X_1,\cdots,X_d){\Phi_0}$ for some polynomial $P$ of degree $n$, which is also guaranteed by the induction hypothesis. The proof is now completed.   
\EndProof\\ 
We will use the spectral theory for commuting self-adjoint operators. The following is sketched in \cite{Xu1,Xu2}. Recall that the self-adjoint operators $T_1,\cdots, T_d$ on a separable Hilbert space with spectral measures $E_1,\cdots,E_d$, respectively, are said to be mutually commuting if their spectral measures commute, i.e., 
\begin{equation}\label{eq:commuting_operators}
E_i(B)E_j(C)=E_j(C)E_i(B),\quad i, j=1,\cdots,d,
\end{equation}
for any Borel sets $B$ and $C$ of $\mathbb R$.  If $T_1,\cdots,T_d$ commute, then 
\begin{equation}\label{eq:spectral_measure}
E:=E_1\otimes\cdots\otimes E_d
\end{equation}
is a spectral measure on $\mathbb R^d$ with values of projections in $\mathcal H$. $E$ is a projection valued measure such that 
\[
E(B_1\times\cdots\times B_d)=E_1(B_1)\cdots E_d(B_d)
\]
for any Borel sets $B_1,\cdots,B_d\subset \mathbb R$. We call $E$ the spectral measure of the commuting operators $T_1,\cdots,T_d$. 

When $T_1,\cdots,T_d$ are bounded the condition \eqref{eq:commuting_operators} is equivalent to $T_iT_j=T_jT_i$, $i,j=1,\cdots,d$. However, if $T_1,\cdots,T_d$ are not bounded it is not the case in general, as the famous example by Nelson shows \cite{RS}. 
 The following spectral theorem which we will use is summarized in \cite{Xu2}. 
\begin{thm}\label{thm:spectral_theory}
Let $\mathcal H$ be a separable Hilbert space and $T_1,\cdots,T_d$ be commuting family of  self-adjoint operators on $\mathcal H$.  If $\Phi_0$ is a cyclic vector in $\mathcal H$ with respect to $T_1,\cdots,T_d$, then $T_1,\cdots,T_d$ are unitarily equivalent to the multiplication operators $M_1,\cdots,M_d$, respectively,
\begin{equation}\label{eq:multiplication_operator}
(M_if)({\bf x})=x_if({\bf x}), \quad 1\le i\le d,
\end{equation}
defined on $L^2(\mathbb R^d,\mu)$, where the measure $\mu$ is defined by $\mu(B)=\langle \Phi_0,E(B)\Phi_0\rangle$. In particular if $\{T_i\}_{i=1}^d$ are bounded operators then $\mu$ is supported on a compact set $S\subset S_1\times\cdots\times S_d$ where $S_i$'s are the spectrum of $T_i$'s, respectively. 
\end{thm}

\Proof[ of Theorem \ref{thm:reconstruction}]  By abuse of the notations let us denote the closures of $\{X_i\}_{i=1}^d$ by the same symbols.  
Therefore  $X_i$'s are mutually commuting self-adjoint operators. By using Lemma \ref{lem:commutative_algebra} it follows from Theorem \ref{thm:spectral_theory} that $X_i$'s are unitarily equivalent to the multiplication operators on $L^2(\mathbb R^d,\mu)$.  in particular, we have 
\begin{equation}\label{eq:measure_representation}
\langle \Phi_0,P(X_1,\cdots,X_d)\Phi_0\rangle=\int P(x_1,\cdots,x_d)d\mu({\bf x}),
\end{equation}
for any polynomial $P(x_1,\cdots, x_d)$. We notice that by Proposition \ref{prop:moments} the interacting Fock space structure defined by this measure $\mu$ is unitarily equivalent to the one that we started with. That is, the reconstruction has been established. 
\EndProof

From now on we give a sufficient condition for (iii) of Theorem \ref{thm:reconstruction} in the case that $X_i$'s are not bounded. We will use a criterion for essential self-adjointness of semibounded operators developed by Jorgensen.
\begin{thm}(\cite[Theorem 1]{J})\label{thm:criterion_ess. s.-a.}
Let $L$ be a semibounded and densely defined operator in a Hilbert space $\mathcal H$. Assume that there is an increasing sequence $\{P_n\}$ of self-adjoint projections in $\mathcal H$ whose supremum is equal to the identity operator such that
\begin{enumerate}
\item[(i)] $\mathrm{ran}\,(P_n)$ is contained in $\mathrm{dom}\,(L)$ for all $n$; 
\item[(ii)] There is a positive integer $k$ such that the range of $LP_n$ is contained in that of $P_{n+k}$ for all $n$; 
\item[(iii)] $\|(I-P_n)LP_n\|\le a_n$ for some sequence $\{a_n\}$ of positive numbers satisfying 
\[
\sum_{n=1}^\infty a_n^{-1/2}=\infty.
\]
\end{enumerate}
Then, the restriction of $L$ to $\cup_n\mathrm{ran}\,(P_n)$ is essentially self-adjont.
\end{thm}
We also need the following lemma which is due to Nelson \cite{N} (cf. \cite{Xu1}). 
\begin{lem}\label{lem:nelson}
Let $T$ and $S$ be symmetric operators in a Hilbert space $\mathcal H$ and let $\mathcal D$ be a dense subspace of $\mathcal H$ such that $\mathcal D$ is contained in the domain of $T^2$, $S^2$, $TS$, and $ST$, and such that $TS\psi=ST\psi$ for all $\psi\in \mathcal D$. If the restriction of $S^2+T^2$ to $\mathcal D$ is essentially self-adjoint then $T$ and $S$ are essentially self-adjoint and $\overline{T}$ and $\overline{S}$ commute, where $\overline{T}$ stands for the closure of $T$.
\end{lem}
\begin{prop}\label{prop:sufficiency_for_(iii)}
Suppose that there is a sequence  $\{a_n\}$ of positive numbers satisfying $\sum_{n=0}^\infty a_n^{-1/2}=\infty$ and  
such that 
\begin{equation}\label{eq:sufficiency}
\|(I-P_{n]})X_i^2P_{n]}\|\le a_n, \quad i=1,\cdots,d.
\end{equation}
Then $\{X_i:i=1,\cdots,d\}$ are essentially self-adjoint and the closures $\{\overline{X}_i:i=1,\cdots,d\}$ are mutually commuting.
\end{prop}
\begin{lem}\label{lem:ess._self_adj.}
Under the hypothesis of Proposition \ref{prop:sufficiency_for_(iii)} the operators $X_i^2$ and their sums $\{X_i^2+X_j^2\}$, $i,j=1,\cdots,d$, are essentially self-adjoint on the domain $\mathcal D$.
\end{lem}
\Proof For each $i=1,\cdots, d$, since $X_i$ is symmetric $X_i^2$ is positive definite on $\mathcal D$. Thus it is bounded from below. Notice that $\cup_{n\ge 0}\mathrm{ran}P_n=\mathcal D$ and $\mathrm{ran}X_i^2P_{n]}\subset \mathrm{ran}P_{n+2]}$. The result now follows from Theorem \ref{thm:criterion_ess. s.-a.}. The same argument applies also to the sums $X_i^2+X_j^2$.
\EndProof\\
\Proof[ of Proposition \ref{prop:sufficiency_for_(iii)}] The proof follows from Lemma \ref{lem:nelson} and Lemma \ref{lem:ess._self_adj.}.
\EndProof

\subsection{Example: univariate Favard's theory and product measures}\label{subsec:example_1-dim_Favard}
In this subsection we discuss one-dimensional theory and product measures. \\
{\bf Univariate Favard's theorem}. First we consider the one-mode interacting Fock space. Let $(\{\omega_n \}, \{\alpha_n \})$ be a Jacobi sequence as in section \ref{sec:preliminary}. Let $\mathcal H_0:=\mathbb C{\bf 1}$ and for $n\ge 1$, let $\mathcal H_n:=\mathbb C^{\widehat\otimes n}$ equipped with the inner product defined by
\begin{equation}\label{eq:1-dim_inner_product}
\langle {\bf e}^{\widehat\otimes n}, {\bf e}^{\widehat\otimes n}\rangle_n:=\prod_{k=1}^n\omega_k
\end{equation}
and by a linear extension. We define $A^+$, $A^0$, and $A^-$ as follows: 
\begin{eqnarray*}
&&\left. A^+\right|_{\mathcal H_n}:{\bf e}^{\widehat\otimes n}\mapsto {\bf e}^{\widehat\otimes (n+1)}\in \mathcal H_{n+1},\\
&&\left. A^0\right|_{\mathcal H_n}:{\bf e}^{\widehat\otimes n}\mapsto \a_{n+1}{\bf e}^{\widehat\otimes n}\in \mathcal H_n,\\
&& \left. A^-\right|_{\mathcal H_n}:{\bf e}^{\widehat\otimes n}\mapsto  \omega_{n}{\bf e}^{\widehat\otimes (n-1)}\in \mathcal H_{n-1}, \quad n\ge 1,\\
&&  \left. A^-\right|_{\mathcal H_0}:=0,
\end{eqnarray*}
and by a linear extension. 
It is promptly checked that $A^-=(A^+)^*$ and the properties (i) and (ii) in the statement of Theorem \ref{thm:reconstruction} are satisfied.  Thus if the operator norms of $\left.A^+\right|_{\mathcal H_n}$ and $\left.A^0\right|_{\mathcal H_n}$ are moderate to satisfy the condition \eqref{eq:sufficiency}, which amounts to saying that the Jacobi sequences do not increase too much fast, then by Theorem \ref{thm:reconstruction} there is a probability measure $\mu$ on $\mathbb R$ such that the sequences $\{\omega_n\}$ and $\{\a_n\}$ are the Jacobi sequences corresponding to the measure $\mu$.\\[1ex]
{\bf Product measures}. Let $\mu_1$ and $\mu_2$ be two probability measures on $\mathbb R$ with Jacobi sequences $(\{\omega_n\},\{\alpha_n\})$ and $(\{\eta_n\},\{\beta_n\})$, respectively. Let $\mathcal H_n$ be the vector space $(\mathbb C^2)^{\widehat \otimes n}$ equipped with an inner product defined as follows: for ${\bf n}=(n_1,n_2), \, {\bf m}=(m_1,m_2)\in \mathcal I_2^{(n)}$, 
\begin{equation}\label{eq:inner_product}
\langle {\bf e}_1^{\widehat\otimes n_1}\widehat\otimes{\bf e}_2^{\widehat\otimes n_2},{\bf e}_1^{\widehat\otimes m_1}\widehat\otimes{\bf e}_2^{\widehat\otimes m_2}\rangle_n:=\delta_{n_1,m_1}\delta_{n_2,m_2}\prod_{k=1}^{n_1}\omega_k\prod_{l=1}^{n_2}\eta_l, 
\end{equation}
and by a linear extension.
Define the creation, preservation, and annihilation operators as follows: for ${\bf n}=(n_1,n_2)\in \mathcal I_2^{(n)}$,
 \begin{eqnarray*}
&&\left. A_1^+\right|_{\mathcal H_n}:{\bf e}_1^{\widehat\otimes n_1}\widehat\otimes{\bf e}_2^{\widehat\otimes n_2}\mapsto  {\bf e}_1^{\widehat\otimes (n_1+1)}\widehat\otimes{\bf e}_2^{\widehat\otimes n_2}\in \mathcal H_{n+1},\\
&&\left. A_2^+\right|_{\mathcal H_n}:{\bf e}_1^{\widehat\otimes n_1}\widehat\otimes{\bf e}_2^{\widehat\otimes n_2}\mapsto  {\bf e}_1^{\widehat\otimes n_1}\widehat\otimes{\bf e}_2^{\widehat\otimes (n_2+1)}\in \mathcal H_{n+1},\\
&&\left. A_1^0\right|_{\mathcal H_n}:{\bf e}_1^{\otimes n_1}\widehat\otimes{\bf e}_2^{\widehat\otimes n_2}\mapsto \a_{n_1+1}{\bf e}_1^{\widehat\otimes n_1}\widehat\otimes{\bf e}_2^{\widehat\otimes n_2}\in \mathcal H_n,\\
&&\left. A_2^0\right|_{\mathcal H_n}:{\bf e}_1^{\widehat\otimes n_1}\widehat\otimes{\bf e}_2^{\widehat\otimes n_2}\mapsto \beta_{n_2+1}{\bf e}_1^{\widehat\otimes n_1}\widehat\otimes{\bf e}_2^{\widehat\otimes n_2}\in \mathcal H_n,\\
&&\left. A_1^-\right|_{\mathcal H_n}:{\bf e}_1^{\widehat\otimes n_1}\widehat\otimes{\bf e}_2^{\widehat\otimes n_2}\mapsto  \omega_{n_1} {\bf e}_1^{\widehat\otimes (n_1-1)}\widehat\otimes{\bf e}_2^{\widehat\otimes n_2}\in \mathcal H_{n-1},\quad n\ge 1,\\
&&\left. A_2^-\right|_{\mathcal H_n}:{\bf e}_1^{\widehat\otimes n_1}\widehat\otimes{\bf e}_2^{\widehat\otimes n_2}\mapsto  \eta_{n_2} {\bf e}_1^{\widehat\otimes n_1}\widehat\otimes{\bf e}_2^{\widehat\otimes (n_2-1)}\in \mathcal H_{n-1},\quad n\ge 1,\\
&&  \left. A_1^-\right|_{\mathcal H_0}:=0, \quad \left. A_2^-\right|_{\mathcal H_0}:=0,
\end{eqnarray*} 
and a linear extension. It is easy to check that $A_i^-=(A_i^+)^*$, $i=1,2$, and the properties (i) and (ii) of Theorem \ref{thm:reconstruction} are satisfied. Thus by Theorem \ref{thm:reconstruction}, if the Jacobi sequences are moderate to satisfy \eqref{eq:sufficiency} there is a probability measure $\mu$ on $\mathbb R^2$ whose interacting Fock space structure reproduces the one given in the above. We can obviously extend the argument to any $d$-dimensional product measures. We will consider other example in subsection \ref{subsec:moments_uniform_circle}.

\section{CAP operators and the form generator}\label{sec:form_generator}
Recall that the pre-Hilbert space $\mathcal H_n$ in the interacting Fock space related to a probability measure on $\mathbb R^d$ is the vector space $(\mathbb C^d)^{\widehat\otimes n}$ equipped with a pre-scalar product $\langle\cdot,\cdot\rangle_n$. As a reference, we also regard $(\mathbb C^d)^{\widehat\otimes n}$ as a Hilbert space equipped with the canonical inner product, which we denote by $(\cdot,\cdot)_0$. We let $\mathcal H_{n,0}:=((\mathbb C^d)^{\widehat\otimes n}, (\cdot,\cdot)_0)$. Since $\langle\cdot,\cdot\rangle_n$ defines a positive definite quadratic form on $\mathcal H_{n,0}$, there is a positive definite operator $\Omega_n:\mathcal H_{n,0}\to \mathcal H_{n,0}$ such that 
\begin{equation}\label{eq:form_generator}
\langle\cdot,\cdot\rangle_n=(\cdot,\Omega_n\cdot)_0.
\end{equation}
From the theory developed before we easily get a matrix representation of $\Omega_n$ by using the creation operators: for ${\bf n}=(n_1,\cdots,n_d)$, ${\bf m}=(m_1,\cdots,m_d)\in \mathcal I^{(n)}$,
\begin{eqnarray}\label{eq:Omega_n_representation}
&&({\bf e}_1^{\widehat\otimes n_1}\widehat \otimes \cdots \widehat \otimes {\bf e}_d^{\widehat\otimes n_d}, \Omega_n {\bf e}_1^{\widehat\otimes m_1}\widehat \otimes \cdots \widehat \otimes {\bf e}_d^{\widehat\otimes m_d})_0\nonumber\\
&=&\langle {\bf e}_1^{\widehat\otimes n_1}\widehat \otimes \cdots \widehat \otimes {\bf e}_d^{\widehat\otimes n_d},   {\bf e}_1^{\widehat\otimes m_1}\widehat \otimes \cdots \widehat \otimes {\bf e}_d^{\widehat\otimes m_d}\rangle_n\nonumber\\
&=&\langle  (A_1^+)^{n_1}\cdots(A_d^+)^{n_d}{\Phi_0},(A_1^+)^{m_1}\cdots(A_d^+)^{m_d}{\Phi_0}\rangle_n.
\end{eqnarray} 
By this we see that given a probability measure on $\mathbb R^d$, we get interacting Fock space $\mathcal H=\oplus_{n=0}^\infty\mathcal H_n$, a sequence of positive definite operators $\Omega_n:\mathcal H_{n,0}\to \mathcal H_{n,0}$, and sequences of Hermitian operators $B_{i|n}:\mathcal H_n\to \mathcal H_n$, $i=1,\cdots,d$, which are in fact defined by $B_{i|n}:=\left. A_i^0\right|_{\mathcal H_n}$. It is worth mentioning that the creation and annihilation operators play a role in the definition of $\Omega_n$ implicitly and moreover, the operators satisfy the conditions (i), (ii), and (iii) in the statement of Theorem  \ref{thm:reconstruction}.

Next let us consider the converse problem. So, suppose that there is a sequence of positive definite operators $\Omega_n:\mathcal H_{n,0}\to \mathcal H_{n,0}$, and  for each $i=1,\cdots, d$, suppose that there is a sequence of operators  $B_{i|n}:\mathcal H_{n,0}\to \mathcal H_{n,0}$. Our purpose is to see under what conditions they would construct a probability measure on $\mathbb R^d$. We proceed in the following steps. 
\begin{enumerate}
\item[(i)] Interacting Fock space. We can define an interacting Fock space as follows. Define a pre-scalar product  $\langle \cdot,\cdot\rangle_n:=(\cdot,\Omega_n\cdot)_0$ and let $\mathcal H_n:=((\mathbb C^d)^{\widehat\otimes n}, \langle \cdot,\cdot\rangle_n)$. The interacting Fock space is denoted by $\mathcal H:=\oplus_{n=0}^\infty \mathcal H_n$.\\[-3ex]
\item[(ii)] Creation and annihilation operators.  As usual we define $A_i^+:\mathcal H\to \mathcal H$, $i=1,\cdots,d$,  by 
\[
\left. A_i^+\right|_{\mathcal H_n}: {\bf e}_1^{\widehat\otimes n_1}\widehat \otimes \cdots \widehat \otimes {\bf e}_d^{\widehat\otimes n_d}\mapsto {\bf e}_1^{\widehat\otimes n_1}\widehat \otimes \cdots \widehat \otimes  {\bf e}_i^{\widehat\otimes (n_i+1)}\widehat \otimes\cdots \widehat \otimes {\bf e}_d^{\widehat\otimes n_d},
\]
and by a linear extension. We let $A_i^-$ the adjoint of $A_i^+$ for $i=1,\cdots, d$, by defining $\left. A_i^-\right|_{\mathcal H_n}:=\left(\left. A_i^+\right|_{\mathcal H_{n-1}}\right)^*$ and $\left. A_i^-\right|_{\mathcal H_0}:=0$.\\[-3ex]
\item[(iii)] Preservation operators. We let $A_i^0:=\oplus_{n=0}^\infty B_{i|n}$, $i=1,\cdots,d$.
\end{enumerate}
We are now ready to state another reconstruction theorem.
\begin{thm}\label{thm:multi_dim_Favard}
Given a probability measure on $\mathbb R^d$, there is a sequence of positive definite operators $\{\Omega_n\}$ satisfying \eqref{eq:form_generator}-\eqref{eq:Omega_n_representation}, and for each $i=1,\cdots, d$, there is a sequence of Hermitian operators $B_{i|n}:\mathcal H_n\to \mathcal H_n$. On the other hand, suppose that we are given a sequence of positive definite operators  $\Omega_n:\mathcal H_{n,0}\to \mathcal H_{n,0}$, and  for each $i=1,\cdots, d$, a sequence of operators  $B_{i|n}:\mathcal H_{n,0}\to \mathcal H_{n,0}$ so that we could construct an interacting Fock space, creation, annihilation, and preservation operators via a process (i)-(iii) above. Suppose that the operators $\left. A_i^0\right|_{\mathcal H_n}$ are Hermitian and that the system of operators $\{A_i^+, \, A_i^0,\,\,A_i^-:i=1,\cdots,d\}$ thus defined satisfy  the conditions (i), (ii), and (iii) in the statement of Theorem  \ref{thm:reconstruction}. Then there is a probability measure $\mu$ on $\mathbb R^d$ such that the operators  $\{\Omega_n\}$ and $\{B_{i|n}\}$ are reconstructed from the measure $\mu$. 
\end{thm}  
\Proof
By  taking $B_{i|n}:=\left. A_i^0\right|_{\mathcal H_n}$, the forward direction was already observed above. The converse follows from Theorem \ref{thm:reconstruction}.
\EndProof  
\begin{ex}\label{ex:diagonal}
In this example let us consider the simplest example for two dimensional space, namely we consider the case where $\Omega_n$'s are diagonal and $B_{i|n}\equiv 0$ for all $n\ge 0$ and $i=1,2$. For ${\bf n}=(n_1,  n_2)$, ${\bf m}=(m_1, m_d)\in \mathcal I_2^{(n)}$, let
\begin{equation}\label{eq:diagonal}
({\bf e}_1^{\widehat\otimes m_1}\widehat\otimes{\bf e}_2^{\widehat\otimes m_2}, \Omega_n {\bf e}_1^{\widehat\otimes n_1}\widehat\otimes{\bf e}_2^{\widehat\otimes n_2})_0:=\delta_{m_1,n_1}\delta_{m_2,n_2}d_{m_1}^{(n)},
\end{equation}
where $d_{k}^{(n)}$'s, $k=0,\cdots,n$,  are positive diagonal components of $\Omega_n$. Recall by step (i) mentioned in this section that the inner product $\langle \cdot, \cdot\rangle_n$ on $(\mathbb C^2)^{\widehat\otimes n}$ is defined by for ${\bf n}=(n_1,  n_2)$, ${\bf m}=(m_1, m_d)\in \mathcal I_2^{(n)}$
\begin{equation}\label{eq:inner_product_diagonal}
\langle {\bf e}_1^{\widehat\otimes m_1}\widehat\otimes{\bf e}_2^{\widehat\otimes m_2},  {\bf e}_1^{\widehat\otimes n_1}\widehat\otimes{\bf e}_2^{\widehat\otimes n_2}\rangle_n:= ({\bf e}_1^{\widehat\otimes m_1}\widehat\otimes{\bf e}_2^{\widehat\otimes m_2}, \Omega_n {\bf e}_1^{\widehat\otimes n_1}\widehat\otimes{\bf e}_2^{\widehat\otimes n_2})_0,
\end{equation}
and it defines the Hilbert space $\mathcal H_n:=((\mathbb C^2)^{\widehat\otimes n},\langle \cdot, \cdot\rangle_n)$. The creation operators $A_i^+$, $i=1,2$, are canonically defined as 
\begin{equation}\label{eq:creations_diagonal}
A_1^+( {\bf e}_1^{\widehat\otimes n_1}\widehat\otimes{\bf e}_2^{\widehat\otimes n_2})= {\bf e}_1^{\widehat\otimes (n_1+1)}\widehat\otimes{\bf e}_2^{\widehat\otimes n_2}, \quad A_2^+( {\bf e}_1^{\widehat\otimes n_1}\widehat\otimes{\bf e}_2^{\widehat\otimes n_2})= {\bf e}_1^{\widehat\otimes n_1}\widehat\otimes{\bf e}_2^{\widehat\otimes (n_2+1)}.
\end{equation}
Then it is easy to check that the annihilation operators, $A_i^-$, $i=1,2$, which are adjoints of $A_i^+$, $i=1,2$, respectively, are defined as
\begin{eqnarray}\label{eq:annihilation11_diagonal}
A_1^-( {\bf e}_1^{\widehat\otimes n_1}\widehat\otimes{\bf e}_2^{\widehat\otimes n_2})&=&\frac{d^{(n)}_{n_1}}{d^{(n-1)}_{n_1-1}}{\bf e}_1^{\widehat\otimes (n_1-1)}\widehat\otimes{\bf e}_2^{\widehat\otimes n_2}, \quad {\bf n}=(n_1,n_2)\in \mathcal I_2^{(n)}, \,\,n_1\ge 1,\\
\label{eq:annihilation12_diagonal}A_1^-(  {\bf e}_2^{\widehat\otimes n })&=&0,\\
\label{eq:annihilation21_diagonal}A_2^-( {\bf e}_1^{\widehat\otimes n_1}\widehat\otimes{\bf e}_2^{\widehat\otimes n_2})&=&\frac{d^{(n)}_{n_1}}{d^{(n-1)}_{n_1}}{\bf e}_1^{\widehat\otimes n_1 }\widehat\otimes{\bf e}_2^{\widehat\otimes (n_2-1)}, \quad {\bf n}=(n_1,n_2)\in \mathcal I_2^{(n)}, \,\,n_2\ge 1,\\
\label{eq:annihilation22_diagonal}A_2^-(  {\bf e}_1^{\widehat\otimes n })&=&0.
\end{eqnarray}
Now the commutation relations \eqref{eq:CR_1} and \eqref{eq:CR_2} are trivially satisfied. In order that the commutation relation \eqref{eq:CR_3} is satisfied, from \eqref{eq:creations_diagonal} to \eqref{eq:annihilation22_diagonal}, the matrix components $d^{(n)}_k$ should satisfy
\begin{equation}\label{eq:diagonal_required}
\frac{d^{(n)}_{n_1}}{d^{(n-1)}_{n_1}}=\frac{d^{(n+1)}_{n_1+1}}{d^{(n)}_{n_1+1}}, \quad \frac{d^{(n+1)}_{n_1}}{d^{(n)}_{n_1-1}}=\frac{d^{(n)}_{n_1}}{d^{(n-1)}_{n_1-1}},\quad n\ge 1, \,\,1\le n_1\le n-1.
\end{equation} 
Let us define 
\begin{equation}\label{eq:diagonal_candidate}
d^{(n)}_{n_1}:=\prod_{k=1}^{n_1}\omega_k\prod_{l=1}^{n-n_1}\eta_l, \quad 1\le n_1\le n-1,
\end{equation}
where $\{\omega_n\}_{n\ge 1}$ and $\{\eta_n\}_{n\ge 1}$ are any sequences of positive numbers. Then one checks easily that the conditions \eqref{eq:diagonal_required} are satisfied. We see that the measure $\mu$ which is reconstructed from $\{\Omega_n\}$ is the product measure $\mu_1\otimes \mu_2$ with Jacobi sequences  $\{\omega_n\}_{n\ge 1}$ and $\{\eta_n\}_{n\ge 1}$, respectively. See subsection \ref{subsec:example_1-dim_Favard}.
\end{ex} 

\section{Examples}\label{sec:examples}
\subsection{Uniform measure on the unit circle}\label{subsec:uniform_on_circle}
Let $\mu$ be a uniform measure on the unit circle $C$ of $xy$-plane. We start by finding a system of orthonormal polynomials for $\mu$. For each $n\ge 0$, let $u_n(x,y)$ and $v_n(x,y)$ be the real- and imaginary-parts of $(x+iy)^n$, respectively: 
\[
(x+iy)^n=u_n(x,y)+iv_n(x,y).
\]
Notice that $u_n(x,y)$ and $v_n(x,y)$ are polynomials of $x$ and $y$ of degree $n$. For $n\geq 1$, we let $p_n(x,y):=\sqrt{2}u_n(x,y)$ and $q_n(x,y):=\sqrt{2}v_n(x,y)$. 
\begin{lem}\label{lem:ons}
$\{1, p_n(x,y), q_n(x,y)\}_{n=1}^\infty$ is an orthonormal system w.r.t. $\mu$.  
\end{lem}
\Proof
By denoting $z=x+iy$, we have for $m,n\geq 0$
\begin{eqnarray*}
\int(x+iy)^m(x+iy)^{-n}d\mu&=&\frac1{2\pi i}\oint_C z^{m-n-1}dz\\
&=&\delta_{m,n}.
\end{eqnarray*}
On the other hand, since $(x+iy)^{m}=u_m(x,y)+iv_m(x,y)$ and $(x+iy)^{-n}=(x-iy)^n=u_n(x,y)-iv_n(x,y)$ on the circle, the above integral is equal to 
\[
\int(u_mu_n+v_mv_n)d\mu+i\int(u_mv_n-v_mu_n)d\mu.
\]
Thus we have
\begin{equation}\label{eq:orthogonality1}
\int(u_mu_n+v_mv_n)d\mu=\delta_{m,n}\,\mathrm{ and }\, \int(u_mv_n-v_mu_n)d\mu=0.
\end{equation}
Similarly we have the relation
\begin{eqnarray*}
&&\int(x+iy)^m(x+iy)^nd\mu=\delta_{m+n,0}\\
&=&\int(u_mu_n-v_mv_n)d\mu+i\int(u_mv_n+v_mu_n)d\mu.
\end{eqnarray*}
Therefore,
\begin{equation}\label{eq:orthogonality2}
\int(u_mu_n-v_mv_n)d\mu=\delta_{m+n,0}\, \mathrm { and }\,\int(u_mv_n+v_mu_n)d\mu=0.
\end{equation}
The result now easily follows from (\ref{eq:orthogonality1}) and (\ref{eq:orthogonality2}).
\EndProof

Recall the gradation spaces $\mathcal P_n=\mathcal P_{n]}\ominus\mathcal P_{n-1]}$. Since we are working on two dimension, the (algebraic) dimension of $\mathcal P_n$ is $n+1$.
\begin{lem}\label{lem:gradation}
For each $n$, the gradation $\mathcal P_n$ has an (algebraic) basis $\{p_n, q_n, (x^2+y^2-1)x^\alpha y^\beta\}$, where $\alpha$ and $\beta$ runs over $\alpha+\beta=n-2$. Moreover, among them the vectors $(x^2+y^2-1)x^\alpha y^\beta$ are zero-norm vectors.
\end{lem}
\Proof
All the polynomials in $\mathcal P_n$ are of degree $n$. The vectors $(x^2+y^2-1)x^\alpha y^\beta$ are linearly independent and obviously of zero-norm vectors. Now by Lemma \ref{lem:ons} the result follows.
\EndProof
\begin{lem}\label{lem:jacobi_relation}
For $n\geq 1$, the relations hold.
\begin{eqnarray*}
xu_n=\frac12(u_{n+1}+u_{n-1}),&& \quad yu_n=\frac12(v_{n+1}-v_{n-1})\\
xv_n=\frac12(v_{n+1}+v_{n-1}), &&\quad yv_n=\frac12(-u_{n+1}+u_{n-1}).
\end{eqnarray*}
\end{lem}
\Proof
From the decomposition $(x+iy)^n=u_n+iv_n$ we have the relations.
\begin{eqnarray*}
(u_{n+1}+iv_{n+1})&=&(x+iy)(u_n+iv_n)\\
&=&(xu_n-yv_n)+i(xv_n+yu_n),\\
(u_{n-1}+iv_{n-1})&=&(x-iy)(u_n+iv_n)\\
&=&(xu_n+yv_n)+i(xv_n-yu_n).
\end{eqnarray*}
Equating the real- and imaginary-parts in the above relations, we easily get the results.
\EndProof

We are now in a position to compute the Jacobi operators. We denote the creation operators by $a^+_x$ and $a^+_y$ and similarly for the annihilation and preservation operators. Notice that $[a^+_x,a^+_y]=0$. We denote the constant function 1 by $\Phi$.
\begin{lem}\label{lem:representation}
For the orthonormal polynomials $p_n(x,y)$ and $q_n(x,y)$ in the gradation $\mathcal P_n$ we have the relation.
\begin{eqnarray*}
p_n(x,y)&=&p_n(a^+_x,a^+_y)\Phi,\\
q_n(x,y)&=&q_n(a^+_x,a^+_y)\Phi.
\end{eqnarray*}
\end{lem}
\Proof
From the commutativity of $a^+_x$ and $a^+_y$ we have the operator expansion:
\[
(a^+_x+ia^+_y)^n=u_n(a^+_x,a^+_y)+iv_n(a^+_x,a^+_y).
\]
Then it is enough to show that
\begin{equation}\label{eq:representation}
(x+iy)^n=(a^+_x+ia^+_y)^n\Phi.
\end{equation}
In order to prove (\ref{eq:representation}) we use induction. Since $(a^+_x+ia^+_y)\Phi=x+iy$, we are done with $n=1$. Assume the relation (\ref{eq:representation}) holds for $n$. Then by Lemma \ref{lem:jacobi_relation} we see that (\ref{eq:representation}) holds also for $n+1$.
\EndProof\\
Let $\{{\bf e}_1,{\bf e}_2\}$ be the canonical basis of $\mathbb C^2$. For each $n\ge 0$, let $\mathcal B^{(n)}:=\{{\bf b}^{(n)}_i:\,i=1,\cdots,n+1\}$ be the canonical basis of $(\mathbb C^2)^{\hat \otimes n}$ consisting of ${\bf e}_1^{\hat\otimes i_1}\hat\otimes{\bf e}_2^{\hat\otimes i_2}$, $i_1+i_2=n$. Notice that the vectors of $\mathcal B^{(n)}$ are orthogonal to each other, but they are not normalized in general. Recall that we denote the inner product and the induced norm of $(\mathbb C^2)^{\hat \otimes n}$ by $(\cdot,\cdot)_0$ and $|\cdot|_0$, respectively.
Let $\Omega_n:=[\omega^{(n)}_{ij}]_{1\le i,j\le n}$ be the representation of $\Omega_n$ w.r.t. $\mathcal B^{(n)}$. Then we have 
\begin{eqnarray}\label{eq:jacobi_components}
\omega^{(n)}_{ij}&=&\frac1{|{\bf b}^{(n)}_{i}|_0^2}({\bf b}^{(n)}_i,\Omega_n{\bf b}^{(n)}_j)_0\nonumber\\
&=&\frac1{|{\bf b}^{(n)}_{i}|_0^2}\langle{U_n\bf b}^{(n)}_i,U_n{\bf b}^{(n)}_j\rangle_\mu.
\end{eqnarray}
As for examples we compute the matrices of $\Omega_1$ and $\Omega_2$, and $\Omega_3$.
We have $\mathcal B^{(1)}=\{{\bf e}_1,{\bf e}_2\}$. Notice that $U_1{\bf e}_1=x=p_1(x,y)/\sqrt{2}$ and $U_1{\bf e}_2=y=q_1(x,y)/\sqrt{2}$. By the formula (\ref{eq:jacobi_components}) we easily get
\begin{equation}\label{eq:Omega_1}
\Omega_1=\frac12\left[\begin{array}{cc} 1&0\\0&1\end{array}\right].
\end{equation}
We see that the eigenvalue of $\Omega_1$ is $1/2$.
To compute $\Omega_2$, let 
\[\mathcal B^{(2)}=\{{\bf b}^{(2)}_1,{\bf b}^{(2)}_2,{\bf b}^{(2)}_3\}=\{{\bf e}_1\hat\otimes{\bf e}_1,{\bf e}_1\hat\otimes{\bf e}_2,{\bf e}_2\hat\otimes{\bf e}_2\}
,
\]
in that order. We compute, for example, $\omega^{(2)}_{22}$. We have $|{\bf b}^{(2)}_2|_0^2=1/2$. By Lemma \ref{lem:jacobi_relation} we have 
\[
U_2 {\bf b}^{(2)}_2=\frac1{2\sqrt{2}}q_2(x,y).
\]
Thus, by (\ref{eq:jacobi_components}) we have $\omega^{(2)}_{22}=1/4$. In this way we can compute all the components of $\Omega_2$ and the result is 
\begin{equation}\label{eq:Omega_2}
\Omega_2=\frac18\left[\begin{array}{ccc}1&0&-1\\0&2&0\\-1&0&1\end{array}\right].
\end{equation}
The eigenvalues of $\Omega_2$ are $\{1/4, 0\}$. Next we let 
\[
\mathcal B^{(3)}=\{{\bf b}^{(3)}_1,{\bf b}^{(3)}_2,{\bf b}^{(3)}_3,{\bf b}^{(3)}_4\}=\{{\bf e}_1\hat\otimes{\bf e}_1\hat\otimes{\bf e}_1,{\bf e}_1\hat\otimes{\bf e}_1\hat\otimes{\bf e}_2,{\bf e}_1\hat\otimes{\bf e}_2\hat\otimes{\bf e}_2,{\bf e}_2\hat\otimes{\bf e}_2\hat\otimes{\bf e}_2\},
\]
in that order. We notice that $|{\bf b}^{(3)}_1|_0^2=|{\bf b}^{(3)}_4|_0^2=1$ and $|{\bf b}^{(3)}_2|_0^2=|{\bf b}^{(3)}_3|_0^2=1/3$. By  Lemma \ref{lem:jacobi_relation} we get 
\begin{eqnarray*}
U_3{\bf b}^{(3)}_1=\frac1{4\sqrt{2}}p_3(x,y), \quad &&U_3{\bf b}^{(3)}_2=\frac1{4\sqrt{2}}q_3(x,y),\\
U_3{\bf b}^{(3)}_3=-\frac1{4\sqrt{2}}p_3(x,y), \quad &&U_3{\bf b}^{(3)}_4=-\frac1{4\sqrt{2}}q_3(x,y).
\end{eqnarray*}
Thus by (\ref{eq:jacobi_components}) we get
\begin{equation}\label{eq:Omega_3}
\Omega_3=\frac1{32}\left[\begin{array}{cccc}1&0&-1&0\\0&3&0&-3\\
-3&0&3&0\\0&-1&0&1\end{array}\right].
\end{equation}
We can compute that the eigenvalues of $\Omega_3$ are \{1/8,0\}.

It turns out that the rank of $\Omega_n$ is $2$. Thus, we don't need to have such a big matrix of size $n+1$ for the representation of $\Omega_n$. Below we find a reduced form of $\Omega_n$.  
\begin{lem}\label{lem:isomorphism}
The isomorphism operator $U_n:(\mathbb C^d)^{\hat\otimes n}\to \mathcal P_n$ is given by 
\[
U_n({\bf e}_1^{\hat\otimes i_1}\hat\otimes\cdots\hat\otimes{\bf e}_d^{\hat\otimes i_d})=x_1^{i_1}\cdots x_d^{i_d}-  P_{n-1]}(x_1^{i_1}\cdots x_d^{i_d}), \quad i_1+\cdots+i_d=n.
\]
\end{lem}
\Proof
We use induction. For $n=1$, 
\[
U_1({\bf e}_i)=a_i^+\Phi=  P_1x_i=x_i-\langle 1,x_i\rangle =x_i- P_{0]}(x_i).
\]
Suppose the statement of the Lemma holds for $n$. Without loss it is enough to check the relation for ${\bf e}_1^{\hat\otimes (i_1+1)}\hat\otimes{\bf e}_2^{\hat\otimes i_2}\cdots\hat\otimes{\bf e}_d^{\hat\otimes i_d}$, $i_1+\cdots+i_d=n$. By the induction assumption,
\begin{eqnarray*}
&&U_{n+1}({\bf e}_1^{\hat\otimes (i_1+1)}\hat\otimes{\bf e}_2^{\hat\otimes i_2}\cdots\hat\otimes{\bf e}_d^{\hat\otimes i_d})\\
&=&a_{x_1}^+(a_{x_1}^+)^{i_1}\cdots(a_{x_d}^+)^{i_d}\Phi\\
&=&a_{x_1}^+U_n({\bf e}_1^{\hat\otimes i_1}\hat\otimes\cdots\hat\otimes{\bf e}_d^{\hat\otimes i_d})\\
&=&  P_{n+1}\left(x_1^{i_1+1}\cdots x_d^{i_d}-x_1  P_{n-1]}(x_1^{i_1}\cdots x_d^{i_d})\right)\\
&=&x_1^{i_1+1}\cdots x_d^{i_d}-  P_{n]}(x_1^{i_1+1}\cdots x_d^{i_d})-  P_{n+1}(x_1  P_{n-1]}(x_1^{i_1}\cdots x_d^{i_d}))\\
&=&x_1^{i_1+1}\cdots x_d^{i_d}-  P_{n]}(x_1^{i_1+1}\cdots x_d^{i_d}),
\end{eqnarray*}
where we have used $  P_{n+1}(x_1  P_{n-1]}(x_1^{i_1}\cdots x_d^{i_d}))=0$  because $x_1  P_{n-1]}(x_1^{i_1}\cdots x_d^{i_d})\in   P_{n]}$ and $  \mathcal P_{n+1}$ is orthogonal to $\mathcal P_{n]}$.
\EndProof\\[2ex]
Recall from Lemma \ref{lem:gradation} that the basis of $\mathcal P_n$ consists of vectors $p_n$, $q_n$, and $r_n$'s, where $r_n$ is any function of the form $(x^2+y^2-1)x^\alpha y^\beta$ with $\alpha+\beta=n-2$. We let $p_n^{\hat{\otimes}}({\bf e}_1,{\bf e}_2)$, $q_n^{\hat{\otimes}}({\bf e}_1,{\bf e}_2)$, and $r_n^{\hat{\otimes}}({\bf e}_1,{\bf e}_2)$ be the unique elements of  $(\mathbb C^2)^{\hat{\otimes}n}$ such that  their image under $U_n$ are $p_n(x,y), \,q_n(x,y)$, and $r_n(x,y)$, respectively. By using Lemma \ref{lem:isomorphism} it is obvious to see how they look like. Indeed, it inherits the form only from the the part of degree-$n$ monomials. For example, for $p_3(x,y)=\sqrt{2}(x^3-3xy^2)$, since $  P_{2]}(\sqrt{2}(x^3-3xy^2))=0$,   $p_3^{\hat{\otimes}}({\bf e}_1,{\bf e}_2)=\sqrt{2}({\bf e}_1\hat\otimes{\bf e}_1\hat\otimes{\bf e}_1-3{\bf e}_1\hat\otimes{\bf e}_2\hat\otimes{\bf e}_2)$. For $r_n(x,y)=(x^2+y^2-1)x^\alpha y^\beta$, $\alpha+\beta=n-2$, since $r_n(x,y)=(x^2+y^2)x^\alpha y^\beta-  P_{n-1]}((x^2+y^2)x^\alpha y^\beta)$, we have  
$r_n^{\hat\otimes}({\bf e}_1, {\bf e}_2)=({\bf e}_1\hat\otimes{\bf e}_1+{\bf e}_2\hat\otimes{\bf e}_2)\hat\otimes {\bf e}_1^{\hat\otimes \alpha}\hat\otimes {\bf e}_2^{\hat\otimes\beta}$. Notice that the part of degree-$n$ monomials of $r_n(x,y)$ is $(x^2+y^2)x^\alpha y^\beta$ and from this the form of $r_n^{\hat\otimes}({\bf e}_1, {\bf e}_2)$ inherits.
\begin{define}
We say that a linearly independent set $\mathcal C^{(n)}=\{{\bf c}_1,\cdots,{\bf c}_k\}$, $1\le k\le n+1$, is closed for $\Omega_n$ if for any vector ${\bf c}_i\in \mathcal C^{(n)}$, $\Omega_n{\bf c}_i$ is a linear combination of the vectors of $\mathcal C^{(n)}$.
\end{define}
Once one has any closed independent subset $\mathcal C^{(n)}$ for $\Omega_n$, then it is enough to represent $\Omega_n$ in the basis of $\mathcal C^{(n)}$. 
\begin{prop}\label{prop:closed_representation}
For each $n\ge 0$, $\mathcal C^{(n)}:=\{p_n^{\hat{\otimes}}({\bf e}_1,{\bf e}_2), q_n^{\hat{\otimes}}({\bf e}_1,{\bf e}_2)\}$ is closed for $\Omega_n$ and $\widetilde\Omega_n$, the representation of $\Omega_n$ in the basis of $\mathcal C^{(n)}$, is given by 
\begin{equation}\label{eq:reduced_matrix}
\widetilde\Omega_n=\left[\begin{array}{cc}\frac1{|p_n^{\hat{\otimes}}({\bf e}_1,{\bf e}_2)|^2}&0\\0&\frac{1}{|q_n^{\hat{\otimes}}({\bf e}_1,{\bf e}_2)|^2}\end{array}\right].
\end{equation}
\end{prop}
\Proof
It is easy to see that $p_n^{\hat{\otimes}}({\bf e}_1,{\bf e}_2)$, $q_n^{\hat{\otimes}}({\bf e}_1,{\bf e}_2)$, and $r_n^{\hat{\otimes}}({\bf e}_1,{\bf e}_2)$'s constitute the orthogonal basis of $(\mathbb C^2)^{\hat\otimes n}$. By definition we see that 
\[
U_n(p_n^{\hat{\otimes}}({\bf e}_1,{\bf e}_2))=p_n(x,y),
\]
and similar relations for $q_n$ and $r_n$'s. 
For any $r_n^{\hat{\otimes}}({\bf e}_1,{\bf e}_2)$, we have
\[
(p_n^{\hat{\otimes}}({\bf e}_1,{\bf e}_2), \Omega_n r_n^{\hat{\otimes}}({\bf e}_1,{\bf e}_2))_0=\langle U_n(p_n^{\hat{\otimes}}({\bf e}_1,{\bf e}_2)),U_n(r_n^{\hat{\otimes}}({\bf e}_1,{\bf e}_2))\rangle_\mu= \langle p_n,r_n\rangle_\mu =0.
\]
Similarly we have $(q_n^{\hat{\otimes}}({\bf e}_1,{\bf e}_2), \Omega_n r_n^{\hat{\otimes}}({\bf e}_1,{\bf e}_2))_0=0$. Thus we see that $\mathcal C^{(n)}$ is closed for $\Omega_n$. The representation (\ref{eq:reduced_matrix}) follows directly from the definition of $\Omega_n$. 
\EndProof

Here are some examples. We have $\mathcal C^{(1)}=\{\sqrt{2}{\bf e}_1, \sqrt{2}{\bf e}_2\}$ and $|\sqrt{2}{\bf e}_1|_0^2=|\sqrt{2}{\bf e}_2|_0^2=2$. Thus 
\begin{equation}\label{eq:Omega_1_tilde}
\widetilde\Omega_1=\frac12\left[\begin{array}{cc}1&0\\0&1\end{array}\right].
\end{equation}
We have $\mathcal C^{(2)}=\{\sqrt{2}({\bf e}_1\hat\otimes{\bf e}_1-{\bf e}_2\hat\otimes{\bf e}_2), 2\sqrt{2}({\bf e}_1\hat\otimes{\bf e}_2)\}$ and 
\[
|\sqrt{2}({\bf e}_1\hat\otimes{\bf e}_1-{\bf e}_2\hat\otimes{\bf e}_2)|_0^2=4=|2\sqrt{2}({\bf e}_1\hat\otimes{\bf e}_2)|_0^2.
\]
Thus,
\begin{equation}\label{eq:Omega_2_tilde}
\widetilde\Omega_2=\frac14\left[\begin{array}{cc}1&0\\0&1\end{array}\right].
\end{equation}
Now for $n=3$, we have $\mathcal C^{(3)}=\{\sqrt{2}({\bf e}_1\hat\otimes{\bf e}_1\hat\otimes{\bf e}_1-3{\bf e}_1\hat\otimes{\bf e}_2\hat\otimes{\bf e}_2), \sqrt{2}(3{\bf e}_1\hat\otimes{\bf e}_1\hat\otimes{\bf e}_2-{\bf e}_2\hat\otimes{\bf e}_2\hat\otimes{\bf e}_2)\}$ and 
\[
|\sqrt{2}({\bf e}_1\hat\otimes{\bf e}_1\hat\otimes{\bf e}_1-3{\bf e}_1\hat\otimes{\bf e}_2\hat\otimes{\bf e}_2)|_0^2=8=|\sqrt{2}(3{\bf e}_1\hat\otimes{\bf e}_1\hat\otimes{\bf e}_2-{\bf e}_2\hat\otimes{\bf e}_2\hat\otimes{\bf e}_2)|_0^2.
\]
Thus,
\begin{equation}\label{eq:Omega_3_tilde}
\widetilde\Omega_3=\frac18\left[\begin{array}{cc}1&0\\0&1\end{array}\right].
\end{equation}
We remark that the non-zero spectrum of $\Omega_n$'s and $\widetilde\Omega_n$'s are equal to each other for $n=1,2,3$. Of course it must be the case for any $n$.

\subsection{Uniform measure on the half circle}
Let $\mu$ be the probability measure uniformly distributed on the half circle on the $xy$-plane; $\{(x,y)\in \mathbb R^2: x^2+y^2=1,\,y\ge 0\}$. Let us find an orthogonal polynomials for this measure. Let $u_n(x,y)$ and $v_n(x,y)$ be the polynomials introduced in the previous subsection, i.e., they satisfy the equation   $(x+iy)^n=u_n(x,y)+iv_n(x,y)$. It turns out that the gradation structure for this measure is very similar to that of the uniform measure on the circle, which we investigated in the previous section. For each $n\ge 1$ let us define the following polynomials.
\begin{eqnarray*}
r_n(x,y)&:=&\begin{cases} u_n(x,y), & n,\,\,\text{odd}\\
 u_{n/2}(x^2-y^2,2xy), &n,\,\,\text{even}\end{cases},
\\
s_n(x,y)&:=&\begin{cases} v_n(x,y), & n\,\,\text{odd}\\
 v_{n/2}(x^2-y^2,2xy), &n,\,\,\text{even}\end{cases}.
\end{eqnarray*}
For each $n\ge 1$, let $\mathcal Q_n:=\{r_n,s_n \}$. We have the following result. 
\begin{lem}\label{lem:orthogonal_system_half_circle} 
For any $n,m\ge 1$, $Q_n\perp Q_m$ if both $n$ and $m$ are even, or both of them are odd. 
\end{lem}
\Proof 
We deal separately with odd and even cases. First observe from the definition that
\begin{eqnarray*}
(x-iy)^n&=&u_n(x,-y)+iv_n(x,-y)\\
&=&u_n(x,y)-iv_n(x,y).
\end{eqnarray*}
From this we get
\begin{equation}\label{eq:y-symmetry}
u_n(x,-y)=u_n(x,y), \quad v_n(x,-y)=-v_n(x,y).
\end{equation}
Similarly we get
\begin{equation}\label{eq:x-symmetry}
u_n(-x,y)=(-1)^nu_n(x,y), \quad v_n(-x,y)=(-1)^{n+1}v_n(x,y).
\end{equation}
Therefore, if $n$ is odd, we get
\begin{equation}\label{eq:x-symmetry_odd}
u_n(-x,y)=-u_n(x,y), \quad v_n(-x,y)=v_n(x,y), \quad (n,\,\text{odd}).
\end{equation}
Let us just show the orthogonality of $r_n$ and $s_m$. 
When $n$ and $m$ are odd, by \eqref{eq:y-symmetry} and \eqref{eq:x-symmetry_odd}, we have $u_n(-x,-y)=-u_n(x,y)$ and $v_m(-x,-y)=-v_m(x,y)$.
That is, the product $u_nv_m$ is symmetric w.r.t. the origin and hence when we integrate out the product $u_nv_m$ over the unit circle, the integral on the upper half circle and the integral on the lower half circle are the same. Thus, 
\begin{eqnarray*}
0&=&\frac1{2\pi}\int_0^{2\pi}u_n(\cos \theta,\sin \theta)v_m(\cos \theta,\sin \theta)d\theta\\
&=&\frac1{\pi}\int_0^{\pi}u_n(\cos \theta,\sin \theta)v_m(\cos \theta,\sin \theta)d\theta\\
&=&\int u_n(x,y)v_m(x,y)d\mu(x,y).
\end{eqnarray*}
Now for each $n,m\ge 1$, we see by change of variables that 
\begin{eqnarray*}
&&\int u_n(x^2-y^2, 2xy)v_m(x^2-y^2,2xy)d\mu(x,y)\\
&=&\frac1{\pi}\int_0^{\pi}u_n(\cos 2\theta,\sin 2\theta)v_m(\cos 2\theta,\sin 2\theta)d\theta\\
&=&\frac1{2\pi}\int_0^{2\pi}u_n(\cos \theta,\sin \theta)v_m(\cos \theta,\sin \theta)d\theta\\
&=&0.
\end{eqnarray*}
So, if $n$ and $m$ are both even, then $\langle r_n, s_m\rangle_\mu=0$. This ends the proof.
\EndProof\\
Now we can state gradation spaces for the measure $\mu$.
\begin{prop}\label{prop:gradation_uniform_half_circle}
For $n\ge 1$, the gradation space $\mathcal P_n$ has a (algebraic) basis $\{p_n,q_n,(x^2+y^2-1)x^\alpha y^\beta:\alpha+\beta=n-2\}$, where $p_n:=r_n-  P_{n-1]}r_n$ and $q_n:=s_n-  P_{n-1]}s_n$.
\end{prop}
\Proof
Note that any vector of the form $(x^2+y^2-1)x^\alpha y^\beta$, $\alpha+\beta=n-2$, is a polynomial of degree $n$ and it is a $\mu$-zero norm vector. $p_n$ and $q_n$ are monomials of degree $n$, and altogether they have full rank for $\mathcal P_n$. We complete the proof by Lemma \ref{lem:orthogonal_system_half_circle}. 
\EndProof\\
From Proposition \ref{prop:gradation_uniform_half_circle} the following holds.
\begin{cor}
For the uniform measure on the half circle, the ranks of $\Omega_n$ are all 2 for $n\ge 1$.
\end{cor} 

\subsection{Moments of uniform measure on the unit circle}\label{subsec:moments_uniform_circle}
In this subsection we revisit the example of uniform measure on the unit circle which we discussed in subsection \ref{subsec:uniform_on_circle}. Here we compute the CAP operators and find a formula for the moments. We have seen that the gradation spaces $\mathcal P_n$ has dimension 2 consisting of orthonormal basis $\{p_n,q_n\}$ for $n\ge 1$. Notice that once the creation operators come from the measure, that is by the relation $A_i^+=U^*a_i^+U$, it is easy to see that $\|\left. A_i^+\right|_{\mathcal H_n}(\xi_n)\|_{n+1}=0$ whenever $\|\xi_n\|_n=0$. Thus it is enough and very convenient if we represent $\left. A_i^+\right|_{\mathcal H_n}:\mathcal H_n\to \mathcal H_{n+1}$ w.r.t. an orthonormal basis, whenever we can  find it easily. Recall the notations
\[
 p_n^{\hat{\otimes}}({\bf e}_1,{\bf e}_2):=U_n^*(p_n),\quad q_n^{\hat{\otimes}}({\bf e}_1,{\bf e}_2):=U_n^*({q_n}).
\]
Then $\mathcal B_n:=\{p_n^{\hat{\otimes}}({\bf e}_1,{\bf e}_2),\, q_n^{\hat{\otimes}}({\bf e}_1,{\bf e}_2)\}$ constitutes an orthonormal basis for $\mathcal H_n$. We have  
\begin{lem}\label{eq:representation_creation_annihilation}
By using the bases $\mathcal B_n$  ($\mathcal B_0:=\{{\bf 1}\}$) above we have the following matrix representation for CAP operators.
\begin{eqnarray*}
&\left.A_1^+\right|_{\mathcal H_n}=\frac12\left[\begin{matrix}1&0\\0&1\end{matrix}\right],\quad n\ge 1,&\left.A_1^+\right|_{\mathcal H_0}=\frac1{\sqrt{2}}\left[\begin{matrix}1\\0\end{matrix}\right],\\
&\left.A_2^+\right|_{\mathcal H_n}=\frac12\left[\begin{matrix}0&-1\\1&0\end{matrix}\right],\quad n\ge 1,&\left.A_2^+\right|_{\mathcal H_0}=\frac1{\sqrt{2}}\left[\begin{matrix}0\\1\end{matrix}\right],\\
&\left.A_1^-\right|_{\mathcal H_n}=\frac12\left[\begin{matrix}1&0\\0&1\end{matrix}\right],\quad n\ge 2,&\left.A_1^-\right|_{\mathcal H_1}=\frac1{\sqrt{2}}\left[\begin{matrix}1&0\end{matrix}\right],\\
&\left.A_2^-\right|_{\mathcal H_n}=\frac12\left[\begin{matrix}0&1\\-1&0\end{matrix}\right],\quad n\ge 2,&\left.A_2^-\right|_{\mathcal H_1}=\frac1{\sqrt{2}}\left[\begin{matrix}0&1\end{matrix}\right],\\
&A_1^0=0, \quad A_2^0=0.&
\end{eqnarray*}
\end{lem}
\Proof
The proof follows easily from Lemma \ref{lem:jacobi_relation}.  
\EndProof\\
As an example let us compute $\int x^2y^{2}d\m(x,y)$:
\begin{eqnarray*}
\int x^2y^{2}d\m(x,y)&=&\langle 1,x^2y^2 1\rangle_\mu\\
&=&\langle {\Phi_0},X_1^2X_2^2{\Phi_0}\rangle_0\\
&=&\langle {\Phi_0},(A_1^++A_1^-)^2(A_2^++A_2^-)^2{\Phi_0}\rangle_0\\
&=&\langle {\Phi_0},(A_1^-A_1^+A_2^-A_2^+ +A_1^-A_1^-A_2^+A_2^+){\Phi_0}\rangle_0\\
&=&\frac14-\frac18=\frac18.
\end{eqnarray*}
In the last line we have used the formula in Lemma \ref{eq:representation_creation_annihilation} and in the line before it, we notice that among all 16 terms there are only two terms that contribute to the integral. By directly computing, we get also $\int x^2y^{2}d\m(x,y)=1/8$. 

\section{Marginals}
In this section we discuss the marginals of a given measure. Let $\mu$ be a probability measure on $\mbR^d$. For any $1\le k<d$, let $\mathcal S=\{i_1,\cdots,i_k\}\sbs\{1,\cdots,d\}$ be a subset. Without loss we may assume $\mathcal S=\{1,\cdots,k\}$. Let $\m^{(\mathcal S)}$ be the marginal of  $\m$ onto $\prod_{i\in \mathcal S}\mbR$. That is, for any Borel set $A\sbs \prod_{i\in \mathcal S}\mbR$, $\m^{(\mathcal S)}(A):=\m(A\times (\prod_{i\notin \mathcal S}\mbR))$. From the general theory developed in sections \ref{sec:multi-dim_IFS}-\ref{sec:form_generator}, it is straightforward how to construct the CAP operators and form generators (operators $\O_n$ in \eqref{eq:form_generator}) for $\m^{(\mathcal S)}$. Let $\mathcal P^{(\mathcal S)}$ be the space of all polynomials of $x_i$ for $i=1,\cdots,k$. Likely we let $\mcP^{(\mathcal S)}_{n]}$ be the space of all polynomials of $x_i$, $i=1,\cdots,k$, of degree less than or equal to $n$. $\mcP^{(\mathcal S)}_n$ denotes the $n$th gradation space:
\[
\mcP^{(\mathcal S)}_n:=\mcP^{(\mathcal S)}_{n]}\ominus \mcP^{(\mathcal S)}_{n-1]}.
\]
As before we let $P^{(\mathcal S)}_{n]}$ and $P^{(\mathcal S)}_{n}$ the projections onto $\mcP^{(\mathcal S)}_{n]}$ and $\mcP^{(\mathcal S)}_{n}$, respectively.
We notice that $\mcP^{(\mathcal S)}_{n]}$ is a subspace of $\mcP_{n]}$ and for any $(n_1,\cdots,n_k)\in \mathcal I_k^{(n)}$, the vector $(a_1^+)^{n_1}\cdots(a_k^+)^{n_k}\Ph=x_1^{n_1}\cdots x_k^{n_k}-P_{n-1}(x_1^{n_1}\cdots x_k^{n_k})$ belongs to $\mathcal P_n$, but it may not equal to $(a_1^{+,\mathcal S})^{n_1}\cdots(a_k^{+,\mathcal S})^{n_k}\Ph=x_1^{n_1}\cdots x_k^{n_k}-P^{(\mathcal S)}_{n-1}(x_1^{n_1}\cdots x_k^{n_k})$, where $a_i^{+,\mathcal S}$'s are creation operators for $\m_\mathcal S$. Now we define CAP operators by 
\begin{eqnarray*}
\left. a_i^{+,\mathcal S}  \right|_{\mcP_n^{(\mathcal S)}}&:=&\mcP^{(\mathcal S)}_{n+1}x_i\mcP^{(\mathcal S)}_{n},\\
\left. a_i^{-,\mathcal S}  \right|_{\mcP_n^{(\mathcal S)}}&:=&\left(\left. a_i^{+,\mathcal S}  \right|_{\mcP_{n-1}^{(\mathcal S)}}\right)^*,\\
\left. a_i^{0,\mathcal S}  \right|_{\mcP_n^{(\mathcal S)}}&:=&x_i-\left. a_i^{+,\mathcal S}  \right|_{\mcP_n^{(\mathcal S)}}-\left. a_i^{-,\mathcal S}  \right|_{\mcP_n^{(\mathcal S)}},
\end{eqnarray*}
for $i=1,\cdots, k$.
These operators enable us to define the form generator $\O_n^{(\mathcal S)}$: for $(n_1,\cdots,n_k)\in \mathcal I_k^{(n)}$ and $(m_1,\cdots,m_k)\in \mathcal I_k^{(n)}$,
\begin{eqnarray}\label{eq:marginal_form_generator}
&\left({\bf e}_1^{\wh \otimes n_1}\wh\otimes \cdots \wh\otimes{\bf e}_k^{\wh\otimes n_k}, \O_n^{(\mathcal S)}{\bf e}_1^{\wh \otimes m_1}\wh\otimes \cdots \wh\otimes{\bf e}_k^{\wh\otimes m_k}\right)_0\nonumber\\
&\quad :=\left\langle (a_1^{+,\mathcal S})^{n_1}\cdots(a_k^{+,\mathcal S})^{n_k}\Ph,(a_1^{+,\mathcal S})^{m_1}\cdots(a_k^{+,\mathcal S})^{m_k}\Ph\right\rangle_\m.
\end{eqnarray}
In the right hand side, the integration w.r.t. $\m$ is equal to the integration w.r.t. $\m^{(\mathcal S)}$ because the integrand is a function of variables $x_i$ for $i\in \mathcal S$. 
Below we consider some examples. \\[2ex]
{\bf Product measures}. Let $\m:=\m_1\otimes\m_2$ on $\mbR^2$ where $\m_1$ and $\m_2$ are one-dimensional measures with Jacobi sequences $(\{\omega_n\},\{\alpha_n\})$ and $(\{\eta_n\},\{\beta_n\})$, respectively. Let $\mathcal S:=\{1\}\sbs\{1,2\}$. Then obviously $\m^{(\mathcal S)}=\m_1$. We will recover this by constructing $\O_n^{(\mathcal S)}$ in \eqref{eq:marginal_form_generator}. Let $\{p_n(x_1)\}$ be the orthogonal polynomials for $\m_1$ satisfying the three-term recurrence relation in \eqref{eq:three-term_recurrence}. By using the fact that $\O_n$'s are diagonal, as noted in Example \ref{ex:diagonal}, we can inductively see that
\[
x_1^n-P_{n-1]}^{(\mathcal S)}(x_1^n)=p_n(x_1).
\]
Therefore, $\o_n^{(\mathcal S)}$, the matrix component of $1\times 1$ matrix $\O_n^{(\mathcal S)}$, is equal to $\langle p_n(x_1),p_n(x_1)\rangle_\m=\prod_{k=1}^n\o_k$. This is the Jacobi coefficients of $\m_1$. \\[2ex]
{\bf Uniform measure on the unit circle}. We come back to the uniform measure on the unit circle discussed in subsection \ref{subsec:uniform_on_circle}. Let $\m$ be the uniform measure on the unit circle and let $\mathcal S:=\{1\}\sbs\{1,2\}$. We want to compute  $\m^{(\mathcal S)}$. Recall the notations in subsection \ref{subsec:uniform_on_circle}:
\[
(x+iy)^n=u_n(x,y)+iv_n(x,y),
\]
and $p_n(x,y)=\sqrt{2}u_n(x,y)$, $q_n(x,y)=\sqrt{2}v_n(x,y)$, which are orthonormal functions for $\m$. The following lemma will be useful.
\begin{lem}\label{lem:op_one-variable}
On the unit circle $x^2+y^2=1$, $u_n(x,y)$ is a polynomial of $x$, say $\wt u_n(x)$, of degree $n$ and the coefficient of the leading term is $2^{n-1}$.
\end{lem}
By directly computing a few number of functions we see that
\begin{eqnarray*}
u_1(x,y)&=&\wt u_1(x)=x,\\
u_2(x,y)&=&\wt u_2(x)=2x^2-1,\\
u_3(x,y)&=&\wt u_3(x)=4x^3-3x,\\
u_4(x,y)&=&\wt u_4(x)=8x^4-8x^2+1,
\end{eqnarray*}
and so on.\\[2ex]
\Proof [ of Lemma \ref{lem:op_one-variable}] On the unit circle, using polar coordinates we get
\[
u_n(x,y)=\text{Re\,}(e^{in\th})=\cos n\th.
\]
Recall an identity for trigonometric functions:
\[
\cos (n+1)\th=2\cos n\th\cos \th-\cos(n-1)\th.
\]
The statement of the lemma is shown by an induction with the above identity. 
\EndProof
\begin{prop}\label{prop:op_marginal_uniform_circle}
For $n\ge 1$ let $p_n^{(\mathcal S)}(x):=2^{-(n-1)}\wt u_n(x)$ and let $p_0^{(\mathcal S)}(x):=1$. Then $\{p_n^{(\mathcal S)}(x):\,n\ge 0\}$ is an orthogonal polynomials for $\m^{(\mathcal S)}$. Moreover, the Jacobi coefficients of $\m^{(\mathcal S)}$ are $\{\o_n\}_{n=1}^\infty=\{1/2,1/4,1/4,\cdots\}$ and $\a_n=0$. Therefore $\m^{(\mathcal S)}$ is the Kesten distribution $\m_{1/2,1/4}$, or an arcsine law with density $\frac{1}{\p}\frac1{\sqrt{1-x^2}}$, $|x|<1$.
\end{prop}
\Proof $u_n(x,y)$ belongs to $\mcP_n$, the $n$th gradation space for the original measure $\m$. Now by Lemma \ref{lem:op_one-variable}, $u_n(x,y)=\wt u_n(x)$ is also a polynomial of the variable $x$ only, thus $\wt u_n(n)$ belongs to $\mcP_n^{(\mathcal S)}$, that is $p_n^{(\mathcal S)}(x):=2^{-(n-1)}\wt u_n(x)=x^n-P_{n-1]}^{(\mathcal S)}(x^n)$, that is $p_n^{(\mathcal S)}(x)=\left(a_1^{+,\mathcal S}\right)^n\Ph$. In order to compute the Jacobi coefficients, we see that for $n\ge 1$
\begin{eqnarray*}
\prod_{k=1}^n\o_k&=&\left\langle \left(a_1^{+,\mathcal S}\right)^n\Ph,\left(a_1^{+,\mathcal S}\right)^n\Ph\right\rangle_{\m_D}\\
&=&\left\langle p_n^{(\mathcal S)}(x)\Ph,p_n^{(\mathcal S)}(x)\Ph\right\rangle_{\m_D}\\
&=&2^{-2(n-1)}\left\langle u_n(x,y),u_n(x,y)\right\rangle_\m\\
&=&2^{-2(n-1)}\frac12.
\end{eqnarray*}
Thus we get $\{\o_n\}_{n=1}^\infty=\{1/2,1/4,1/4,\cdots\}$. 
\EndProof \\[2ex]
{\bf A non-symmetric measure}. Let  $\m:=\frac14(\d_{(2,0)}+\d_{(1,1)}+\d_{(0,0)}+\d_{(1,-1)})$, a point mass on $\mbR^2$. We notice that $\mu$ is the rotation of the product measure $\nu_1\otimes \nu_2$, where $\n_1=\n_2=\frac12(\d_{1/\sqrt{2}}+\d_{-1/\sqrt{2}})$, followed by a translation by 1 in the $x$-axis. Since the orthonormal system of $\nu_1\otimes \nu_2$ is $\{1,\sqrt{2}, \sqrt{2}, 2xy\}$, the orthonormal system of $\m$ is $\{1, (x-1)-y, (x-1)+y, (x-1)^2-y^2\}$. Since we aim at the $x$-marginal, we may rewrite the orthogonal system in the following way. For the degree 1 polynomials we use linear combinations and for the degree 2 polynomial we use the identity $(x-1)^2+y^2=1$ which holds for $\m$-a.e.. Thus we have another orthonormal system for $\m$ of the form
\[
\{1, \sqrt{2}(x-1), \sqrt{2}y, 2(x-1)^2-1\}.
\]
Thus, the orthonormal polynomials for $\m^{(1)}$, the $x$-marginal of $\m$, are 
\[
\{1, \sqrt{2}(x-1), 2(x-1)^2-1\}.
\]
Thus the monic bases \cite{ABD} for $\m$ and $\m^{(1)}$ are 
\begin{equation}\label{eq:monic basis}
\left\{1,  (x-1),  y,  (x-1)^2-\frac12\right\}\text{ and } \left\{1,  (x-1),  (x-1)^2-\frac12\right\},
\end{equation}
respectively. From \eqref{eq:monic basis} we easily compute the Jacobi operators as follows.
\[
\O_0=1,\quad \O_1=\left[\begin{matrix}\frac12&0\\0&\frac12\end{matrix}\right], \quad \O_2=\left[\begin{matrix}\frac14&0&-\frac14\\0&0&0\\-\frac14&0&\frac14\end{matrix}\right]
\]
and 
\[
\O_0^{(1)}=1,\quad \O_1^{(1)}=\frac12,\quad \O_2^{(1)}=\frac14.
\]
We promptly see that $\m^{(1)}=\frac14\d_0+\frac12\d_1+\frac14\d_2$ and the Jacobi sequences are $\{\o_n\}=\{1/2,1/2,0,\cdots\}$ and $\{\a_n\}=\{1,1,1,0,\cdots\}$.
\section{Deficiency rank of Jacobi operator and support of the measure}\label{sec:deficiency_rank}
In the examples of section \ref{sec:examples}, we see that the rank of $\Omega_n$ is $2$ for all $n\ge 1$, i.e., it is uniformly bounded by a constant, or at least, it is less than $d_n$, the possible full rank of $\Omega_n$. Below we discuss this phenomenon. On $\mathbb R^d$, we say that a subset $S\subset\mathbb R^d$ is an algebraic level surface if there is a polynomial $p$ such that $S=\{(x_1,\cdots,x_d)\in \mathbb R^d:\,p(x_1,\cdots,x_d)=0\}$.
\begin{define}
Let $\mu$ be a probability measure on $\mathbb R^d$ with finite moments of all orders. 
By defining $\rho_n:=\text{rank}\,\Omega_n$, we call $\rho:=(\rho_n)_{n\ge 1}$ the rank sequence of $\mu$. We say that $\mu$ has deficiency rank if there is  $n_0$ such that   $\text{rank}\,\Omega_{n_0}$ is strictly less than $d_{n_0}={n_0+d-1\choose d-1}$, the possible maximum rank of $\Omega_{n_0}$. 
\end{define} 
Notice that once $\text{rank}\,\Omega_{n_0}<d_{n_0}$, it is the case for all $n\ge n_0$. 
\begin{thm}\label{thm:deficiency_rank} Let $\mu$ be a probability measure on $\mathbb R^d$ with finite moments of all orders. Then $\mu$ has deficiency rank if and only if the measure $\mu$ is supported on an algebraic level surface.
\end{thm}
\Proof
Suppose that $\mu$ has deficiency rank. Then $\Omega_n$ has an eigenvalue 0 with corresponding eigenvector, say $\xi\in (\mathbb C^d)^{\widehat\otimes n}$. Let $p:=U_n(\xi)\in \mathcal P_n$. Then,
\begin{eqnarray}\label{eq:deficiency_rank}
\int |p|^2d\mu&=&\langle p,p\rangle_\mu\nonumber\\
&=&\langle \xi,\xi\rangle_n\nonumber\\
&=&(\xi,\Omega_n\xi)_0\\ \nonumber
&=&0.
\end{eqnarray}  
This means that $p=0$ $\mu$-a.e. Therefore $\mu$ is supported on the algebraic level surface $\{p=0\}$. Conversely, suppose that $\mu$ is supported on an algebraic level surface, say $\{p=0\}$, for a polynomial $p$ of degree $n$. We may assume $p\in \mathcal P_n$. Let $\xi=U^*(p)\in \mathcal H_n$. By the equality \eqref{eq:deficiency_rank} we have
\[
0=\int |p|^2d\mu=(\xi,\Omega_n\xi)_0.
\]
Thus $\sqrt{\Omega_n}\xi=0$, and hence $\Omega_n\xi=\sqrt{\Omega_n}^2\xi=0$, i.e., $\Omega_n$ has a zero eigenvalue and therefore $\mathrm{rank}\,\Omega_n<d_n$. This ends the proof.  
\EndProof
\begin{rem}
(1) When $\mu$ has deficiency rank, the ranks of $\Omega_n$ may be uniformly bounded by a constant or may increase monotonically. The case of uniform measure on the unit circle   is an example of uniform bound. Now consider the measure $d\mu(x,y)d\pi(z)$ on $\mathbb R^3$. Here $d\mu(x,y)$ is the uniform measure on the unit circle on the $xy$-plane and $d\pi(z)$ is a measure of infinite orthogonal polynomials, e.g., a Gaussian measure on the $z$-axis. Let  $\{p_n(x,y), q_n(x,y)\}_{n\ge 0}$ be the orthogonal systems for $\mu(dxdy)$ as above and let $\{r_n(z)\}_{n\ge 0}$ be the orthogonal polynomials for $\pi(dz)$. Then the orthogonal system for $\mathcal P_n$ is $\{p_k(x,y)r_{n-k}(z)\}_{k=0}^n\cup\{q_k(x,y)r_{n-k}(z)\}_{k=0}^n$. So, the rank of $\Omega_n$ is $2(2n+1)$ which is less than $d_n={n+2\choose 2}=(n+2)(n+1)/2$, and hence the measure $d\mu(x,y)d\pi(z)$ has deficiency rank, but the ranks increase to infinity. Notice that the cylinder $\{x^2+y^2=1\}$ is an algebraic level surface on $\mathbb R^3$. \\
(2) Although the ranks may increase to infinity for a measure of deficiency rank, the increase is negligible in the sense that $\lim_{n\to \infty}\mathrm{rank}\,\Omega_n/d_n=0$. In fact, suppose that $\mu$ is supported on an algebraic level surface $p=0$, where $p$ is a polynomial of degree $k$. Then the dimension of null space of $\mathcal P_n$ is at least $n-k+d-1\choose d-1$ for $n\ge k$. Thus $\lim_{n\to \infty}\mathrm{rank}\,\Omega_n/d_n=0$ since $\lim_{n\to \infty}{n-k+d-1\choose d-1}/d_n=1$.
\end{rem}
\begin{ex}
(1) Any measure with finitely many point masses has deficiency rank because there always exists an algebraic level surface that contains all the mass points.\\
(2) Let $\mu_1$ be a discrete measure with support the natural numbers, e.g., let $\mu_1=\sum_{n=1}^\infty \frac1{2^n}\delta_n$. Let $\mu_2$ be a copy of $\mu_1$ and let $d\mu(x,y):=d\mu_1(x)d\mu_2(y)$ be the product measure. We see that the support of $\mu$ is the two-dimensional lattice points on the first quadrant. Now there is no algebraic level curve that contains all the lattice points on the first quadrant. In fact, let $\{p_n(x)\}_{n\ge 0}$ be the orthogonal polynomials for $\mu_1$. Then an orthogonal system for $\mathcal P_n$ for $\mu$ is $\{p_k(x)p_{n-k}(y)\}_{k=0}^n$, which is of dimension $n+1=d_n$. Thus $\mu$ is not a measure of deficiency rank. \\
(3) Let $d\mu(x,y)$ be the image measure on the curve $y=\sin x$ of the Gaussian measure on the $x$-axis by the map $x\mapsto (x,\sin x)$. Notice that $\mu$ is singular w.r.t. Lebesgue measure on the plane (it lives on a curve). But $\mu$ is not a measure of deficiency rank, because any polynomial of $x$ and $y$ can't be a zero function on the curve $y=\sin x$.
\end{ex}

Let us now more closely look at the relation between the deficiency rank and support of the measure. As we have seen in Theorem \ref{thm:deficiency_rank}, if a measure $\mu$ has deficiency rank then there exist polynomials with zero norm. Let $\mu$ be a probability measure on $\mathbb R^d$ which has finite moments of all orders. Suppose that $\mu$ has deficiency rank and let 
\[
\mathcal N\equiv \mathcal N_\mu:=\{p\in \mathcal P: \int |p|^2d\mu=0\}.
\]
We call $p\in \mathcal N$ a {\it base null polynomial} (simply base) if $p$ has no more factor of zero norm, i.e., there is no pair $h\in \mathcal N$ and $p_1\in \mathcal P$ such that degree of $p_1$ is greater than or equal to 1 and $p=hp_1$. When $p$ is a base, the level surface $\{p=0\}$ we call a {\it base level surface}. 

We recall some of basic facts for polynomial algebra. 
\begin{prop}\label{prop:finitely_generated}
Any ideal of the polynomial algebra $\mathcal P$ is finitely generated.
\end{prop}
The above fundamental result on polynomial rings
traces back to Hilbert.
In fact, the coefficients of our polynomials 
are taken from the real number field which is Noetherian,
so is the polynomial.

We also recall the following

\begin{prop}\label{prop:factorization_ring}
The polynomial algebra $\mathcal{P}$ is a unique factorization ring.
\end{prop}

Hence for the null kernel $\mathcal{N}$ there exist
a finite number of polynomials $f_1,\dots,f_k$ such that
\begin{equation}\label{eqn:N is finitely generated}
\mathcal{N}=\sum_{i=1}^k f_i \mathcal{P},
\end{equation}
where $f_1,\dots,f_k$ are linearly independent.
The algebraic set corresponding to $\mathcal{N}$ is defined by
\begin{equation}\label{eq:support}
S(\mathcal{N})=\bigcap_{i=1}^k \{f_i=0\},
\end{equation}
where 
\[
\{f_i=0\}
=\{(x_1,x_2,\dots,x_d)\in\mathbb{R}^d\,:\,f_i(x_1,x_2,\dots,x_d)=0\}.
\]

The support of $\mu$ is a closed subset of $\mathbb{R}^d$ defined by
\[
\mathrm{supp \,}\mu
=\mathbb{R}^d \Big\backslash \bigcup \{U:\, 
\text{an open set in $\mathbb{R}^d$ such that $\mu(U)=0$}\}
\]
The following is a fundamental relation between deficiency rank and support of the measure.
\begin{thm}\label{thm:deficiency_rank_base_surface}
Let $\mu$ be a probability measure on $\mathbb R^d$ with finite moments of all orders. If $\mu$ has deficiency rank then $\mathrm{supp \,}\mu\subset S(\mathcal{N})$.
\end{thm}
\Proof
It follows easily from Theorem  \ref{thm:deficiency_rank}.
\EndProof
\begin{ex}
(1) Let $\mu$ be the uniform measure on the unit circle or the uniform measure on the half circle of $\mathbb R^2$. We have seen that in both cases the polynomial $x^2+y^2-1$ is the unique base null polynomial. Thus in two cases the measures are supported on this base level surface; $x^2+y^2=1$.\\
(2) On $\mathbb R^2$, let $\mu=\frac14(\delta_{(1,1)}+\delta_{(-1,1)}+\delta_{(-1,-1)}+\delta_{(1,-1)})$. It is not hard to see that the base null polynomials are $x^2-1$, $y^2-1$, $x^2-y^2$, and $x^2+y^2-2$. (the last two are the linear combinations of the first two). Now the intersection of the base level surfaces are exactly four points $\{(1,1), (-1,1), (-1,-1), (1,-1)\}$, the support of the measure. In general, let $\{{\bf a}_i=(a^i_1,\cdots,a^i_d)\in \mathbb R^d:i=1,\cdots,k\}$ be a finite subset and let $\mu:=\sum_{i=1}^km_i\delta_{{\bf a}_i}$ with $\sum_{i=1}^km_i=1$. Then for each $j=1,\cdots,d$, the polynomial $\prod_{i=1}^k(x_j-a^i_j)$ is a base null polynomial. The intersection of the base level surfaces $\prod_{i=1}^k(x_j-a^i_j)=0$, $j=1,\cdots,d$, gives rise to the support of the measure.
\end{ex}

\vskip 1true cm
\noindent\textbf{Acknowledgments.} A. Dhahri acknowledges support by Basic Science Research Program through the National Research Foundation of Korea (NRF) funded by the Ministry of Education (grant 2016R1C1B1010008). The research by H. J. Yoo was supported by Basic Science Research Program through the National
Research Foundation of Korea (NRF) funded by the Ministry of Education (NRF-2016R1D1A1B03936006).

\end{document}